\documentclass[aps,
prd,
twocolumn,
nofootinbib,
superscriptaddress,
print
]{revtex4-2}

\usepackage[utf8]{inputenc}
\usepackage{amsmath,amsfonts,amssymb,mathrsfs}
\usepackage{graphicx}
\usepackage{color}
\usepackage{booktabs}
\usepackage{multirow}
\usepackage{makecell}
\usepackage{siunitx}
\usepackage{ragged2e}
\usepackage{rotating}
\usepackage{float}

\usepackage[dvipsnames]{xcolor}
\definecolor{darkred}{rgb}{0.5,0,0}
\definecolor{darkblue}{rgb}{0,0,0.5}
\definecolor{firebrick}{rgb}{0.75,0.125,0.125}
\definecolor{darkgreen}{rgb}{0,0.5,0}

\usepackage[colorlinks=true,breaklinks=true,
            urlcolor=darkblue,
            citecolor=darkgreen,
            linkcolor=firebrick]{hyperref}

\usepackage{orcidlink}

\usepackage[normalem]{ulem} 
\usepackage{enumitem}
\usepackage[capitalise]{cleveref}
\usepackage{pifont}
\usepackage{subfigure} 
\usepackage{comment}

\long\def\exclude#1{}

\graphicspath{{figures/}}
\begin{document}
\include{jdefs}

\title{Mapping the multimessenger horizon for $\gamma$-ray counterparts of cosmic neutrinos}

\author{Tianyi Ding}
\email{dingt2@unlv.nevada.edu}
\affiliation{Department of Physics and Astronomy, University of Nevada, Las Vegas, NV 89514}
\affiliation{Nevada Center for Astrophysics, University of Nevada, Las Vegas, NV 89514}
\author{Jose Alonso Carpio}
\affiliation{Department of Physics and Astronomy, University of Nevada, Las Vegas, NV 89514}
\affiliation{Nevada Center for Astrophysics, University of Nevada, Las Vegas, NV 89514}
\affiliation{Niels Bohr Institute, University of Copenhagen, Blegdamsvej 17, 2100 Copenhagen, Denmark}
\email{jose.dumler@nbi.ku.dk}
\author{Ali Kheirandish}
\email{ali.kheirandish@unlv.edu}
\affiliation{Department of Physics and Astronomy, University of Nevada, Las Vegas, NV 89514}
\affiliation{Nevada Center for Astrophysics, University of Nevada, Las Vegas, NV 89514}
\author{Kohta Murase}
\email{murase@psu.edu}
\affiliation{Department of Physics, Department of Astronomy, Institute for Gravitation and the Cosmos, The Pennsylvania State University, University Park, PA 16802}
\affiliation{Department of Physics, The University of Tokyo, 7-3-1 Hongo, Bunkyo, Tokyo 113-0033, Japan}
\affiliation{Center for Gravitational Physics and Quantum Information, Yukawa Institute for Theoretical Physics, Kyoto, Kyoto 606-8502 Japan}

\date{\today}

\begin{abstract}
The joint detection of high-energy neutrinos and $\gamma$-rays offers a powerful probe of the most energetic astrophysical accelerators, linking hadronic interactions to observable signatures across cosmic distances. Understanding the conditions under which these two messengers can be simultaneously observed is essential for identifying and characterizing the sources of cosmic neutrinos. In this work, we examine the detectability of $\gamma$-ray cascades associated with neutrino production in source environments that are gamma-ray transparent, focusing on how electromagnetic cascades evolve as $\gamma$-rays propagate through intergalactic space. By modeling photon interactions with the extragalactic background light and deflections in intergalactic magnetic fields, we determine the minimum neutrino fluence for transient emission and neutrino flux for steady emission required to produce a $\gamma$-ray counterpart detectable by $\gamma$-ray observatories, such as \textit{Fermi}-LAT, HAWC, or LHAASO. This defines a “multimessenger horizon” — the region in flux–distance parameter space where neutrinos and $\gamma$-rays are jointly observable. We consider both energy-binned injection tied to IceCube’s sensitivity and a continuous power-law spectrum extending from TeV to PeV energies. Our results reveal that detectability depends strongly on redshift, magnetic field strength, and photon energy, with higher-energy cascades suffering greater attenuation inside filamentary structures. The framework provides new constraints on source opacity, baryon loading, and neutrino production mechanisms, offering a predictive tool for interpreting current events and guiding future multimessenger campaigns.
\end{abstract}

\maketitle

\section{Introduction}
\label{sec:intro}
The detection of high-energy neutrinos of astrophysical origin~\cite{IceCube:2013cdw,IceCube:2013low} marks a transformative moment in our quest to understand the most energetic phenomena in the universe. Multimessenger astronomy, combining information carried by neutrinos, photons, cosmic rays, and gravitational waves, offers a more complete picture of cosmic accelerators by probing different aspects of particle interactions and source environments. Among these messengers, neutrinos play a uniquely powerful role. Unlike photons, which can be absorbed or deflected by intervening matter and radiation, neutrinos travel virtually unimpeded from their production sites, carrying direct information about hadronic processes in extreme astrophysical environments~\citep{Meszaros:2019xej,Halzen:2022pez,Kurahashi:2022utm}.

While $\gamma$-ray observations have been instrumental in identifying energetic astrophysical sources, they alone cannot unambiguously reveal the nature of the underlying particle acceleration mechanisms. In particular, $\gamma$-rays can be produced through both hadronic and leptonic processes, see e.g.,~\cite{Tavecchio:1998xw, Finke:2008pe,Reimer:2011zz,Boettcher:2013wxa}. In hadronic scenarios, accelerated protons (cosmic rays) interact with ambient photons or matter, generating charged and neutral pions that decay into neutrinos and $\gamma$-rays, respectively. 
In contrast, leptonic models involve relativistic electrons that produce $\gamma$-rays via inverse-Compton (IC) scattering or synchrotron emission, but do not generate neutrinos. This overlap in observational signatures creates a fundamental degeneracy: the presence of high-energy $\gamma$-rays does not necessarily imply hadronic acceleration. Consequently, detecting high-energy neutrinos becomes critical as a smoking gun. Confirming a $\gamma$-ray–neutrino coincidence can break this degeneracy, offering definitive evidence of cosmic-ray protons or nuclei~\cite{Halzen:2002pg}. Conversely, the absence of $\gamma$-rays in known neutrino-emitting sources raises further insight into $\gamma$-ray absorption and dense environments~\cite{Murase:2015xka}. 

The observation of a neutrino with $\sim$ 290 TeV energy by the IceCube Neutrino Observatory on September 22, 2017, coincident with a bright $\gamma$-ray flare from the blazar TXS~0506+056~\cite{IceCube:2018dnn,Keivani:2018rnh}, provided an important motivation for multimessenger astronomy with neutrinos. 
This event provided the first compelling evidence linking a cosmic neutrino to an extragalactic source and demonstrated the potential of multimessenger observations to uncover the mechanisms of particle acceleration in active galactic nuclei (AGN). Further analysis revealed a prior neutrino flare from the same source in 2014–2015 with about $13\pm5$ events in 110 days~\cite{IceCube:2018cha}, this time lacking a $\gamma$-ray counterpart~\cite{Padovani:2018acg,Fermi-LAT:2019hte,Petropoulou:2019zqp}. These contrasting observations support the presence of hadronic acceleration, particularly via pion decay in $p\gamma$ interactions in dense photon fields~\cite{Halzen:2018iak, RosalesdeLeon:2021qth}.

\begin{figure}[t]
    \centering
    \includegraphics[width=0.92\columnwidth]{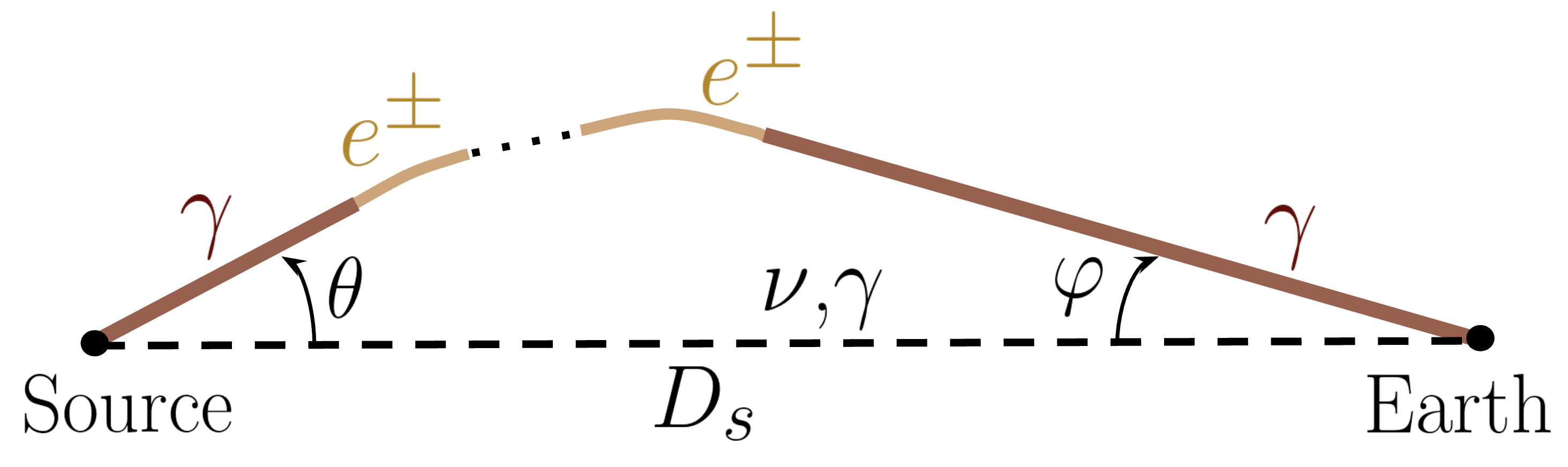}
    \caption{Schematic illustration of the electromagnetic cascade geometry from a source located at a luminosity distance $D_s$ from Earth. A high-energy $\gamma$-ray is emitted at an emission angle \(\theta\) relative to the observer's line of sight and undergoes pair production, generating an electron–positron pair (\(e^\pm\)). These charged particles are deflected by the IGMF and subsequently upscatter background photons via IC scattering, producing secondary  $\gamma$-rays. The final cascade photon arrives at the observer with a detection angle \(\varphi\). Neutrinos and any unattenuated $\gamma$-rays travel along the direct path of luminosity distance \(D_s\). The developed cascade travels a larger distance than neutrinos and unattenuated $\gamma$-rays, which causes a time delay at detection.}
    \label{fig: geo}
\end{figure}

While neutrinos escape their sources and propagate through intergalactic space largely unimpeded, the same is not true for $\gamma$-rays. High-energy $\gamma$-rays produced alongside neutrinos may undergo attenuation due to interactions with dense photon or matter fields inside the source, namely via $\gamma\gamma$ or $\gamma A$ absorption, rendering the source opaque to its own high-energy emission~\cite{Aharonian:2008su}. Beyond the source, $\gamma$-rays traveling cosmological distances interact with photons of the extragalactic background light (EBL) and the cosmic microwave background (CMB), initiating electromagnetic cascades through pair production~\cite{Nikishov:1962rmq,Gould:1967zza,PhysRev.155.1404} and IC scattering. 
These cascades generate lower-energy secondary $\gamma$-rays, which are delayed and deflected by the intergalactic magnetic field (IGMF). The delayed $\gamma$-ray emission is called ``pair echo''~\cite{Plaga:1995ins,Murase:2008pe,Takahashi:2008pc,Murase:2008yy}, whereas the extended $\gamma$-ray emission is called ``pair halo''~\cite{Aharonian:1993vz,Neronov:2006lki,Neronov:2009gh}. 
The $\gamma$-ray propagation process is illustrated in Figure~\ref{fig: geo}. 

Murase et al.~\cite{Murase:2012df} estimated detection horizons for high-energy neutrinos and cascade $\gamma$-rays (see their section~4), and calculated the number of sources within these. However, the detectability of cascade $\gamma$-ray sources depends sensitively on the IGMF strength, source redshift, and the angular resolution of the instruments. As the cascade is significantly extended or delayed, the resulting $\gamma$-ray emission may arrive far outside the observation window or the field of view in multimessenger searches, causing it to evade detection or to no longer be recognized as originating from the same astrophysical event as the neutrino. These effects collectively impose a fundamental limitation on our ability to associate $\gamma$-ray counterparts with neutrino sources, motivating the need to map the conditions under which such multimessenger associations are feasible.

In this study, we systematically investigate the detectability of $\gamma$-ray counterparts to high-energy neutrino sources. In particular, we simulate the propagation of $\gamma$-rays accompanying neutrinos in photohadronic ($p\gamma$) interactions, considering IceCube’s point-source sensitivity threshold. 
For this purpose, we implement detailed Monte Carlo simulations with \texttt{CRPropa3}~\cite{AlvesBatista:2016vpy} to model the development of electromagnetic cascades as $\gamma$-rays traverse structured regions (filaments) and voids with varying photon and magnetic field environments. 
In section~\ref{sec:MMinterface}, we discuss how the initial $\gamma$-ray flux is inferred from the neutrino signal, and how internal and extragalactic attenuation affect its detectability. Section~\ref{sec:Method} then establishes criteria for transient and steady $\gamma$-ray emissions and evaluates their observational prospects against the sensitivity of instruments such as \textit{Fermi}-LAT, LHAASO, and HAWC. Finally, in Section~\ref{sec:Results}, we present the resulting multimessenger horizon, identifying the redshift and flux ranges where $\gamma$-ray counterparts are expected to be observable.

Throughout the work, a standard $\Lambda$CDM cosmology is assumed, with matter density $\Omega_\mathrm{m} = 0.3$, vacuum energy density $\Omega_\Lambda = 0.7$, and Hubble constant $H_0 = 70 \, \text{km s}^{-1} \, 
\text{Mpc}^{-1}$~\cite{Planck:2018vyg}. 

\section{Simulation Setup}
\label{sec: Simulation_intro}

High-energy $\gamma$-rays that leave the sources will interact with the EBL or CMB and thus produce electron–positron pairs. These pairs subsequently go through IC scattering with CMB photons, generating secondary $\gamma$-rays at lower energies and thereby initiating an electromagnetic cascade. 
The presence of IGMFs can deflect the electron–positron pairs, elongating the travel path and introducing a time delay between the arrival of the secondary $\gamma$-rays and the primary $\gamma$-rays that could originate from neutral pions produced simultaneously with high-energy neutrinos. 

The vast majority of galaxies are embedded in the cosmic web, primarily within filaments and clusters~\cite{Poudel:2016ghk, Constantin:2007ab}. 
These structures, characterized by enhanced gas density and magnetic field strengths relative to surrounding voids, should influence $\gamma$-ray propagation and their observability significantly~\cite{Murase:2011yw,Murase:2011cy,Oikonomou:2014xea,Takami:2014zva,Das:2025zur,Sherman:2025gir}. 
Therefore, to systematically investigate these propagation effects, we first consider the propagation of $\gamma$-rays within cosmic filaments, regions characterized by relatively strong magnetic fields. Subsequently, we examine the propagation of $\gamma$-rays as they emerge into cosmological voids, which feature extremely weak magnetic fields.

To accurately examine the spectral, angular, and temporal signatures of electromagnetic cascades in both filament and void regions, we conduct detailed three-dimensional Monte Carlo simulations using \texttt{CRPropa3}~\cite{AlvesBatista:2016vpy}. \texttt{CRPropa3} is an open-source numerical framework specifically designed for simulating the propagation of ultra-high-energy cosmic rays, $\gamma$-rays, and neutrinos through astrophysical environments on cosmological scales. To ensure precise accounting of redshift-dependent interactions and trajectory lengths, particularly for sources at high redshifts, we implemented appropriate modifications within the \texttt{CRPropa3} framework (see appendix for details).

\subsection{Background Photon Fields}

\begin{figure}[t]
    \centering
    \includegraphics[width=1\columnwidth]{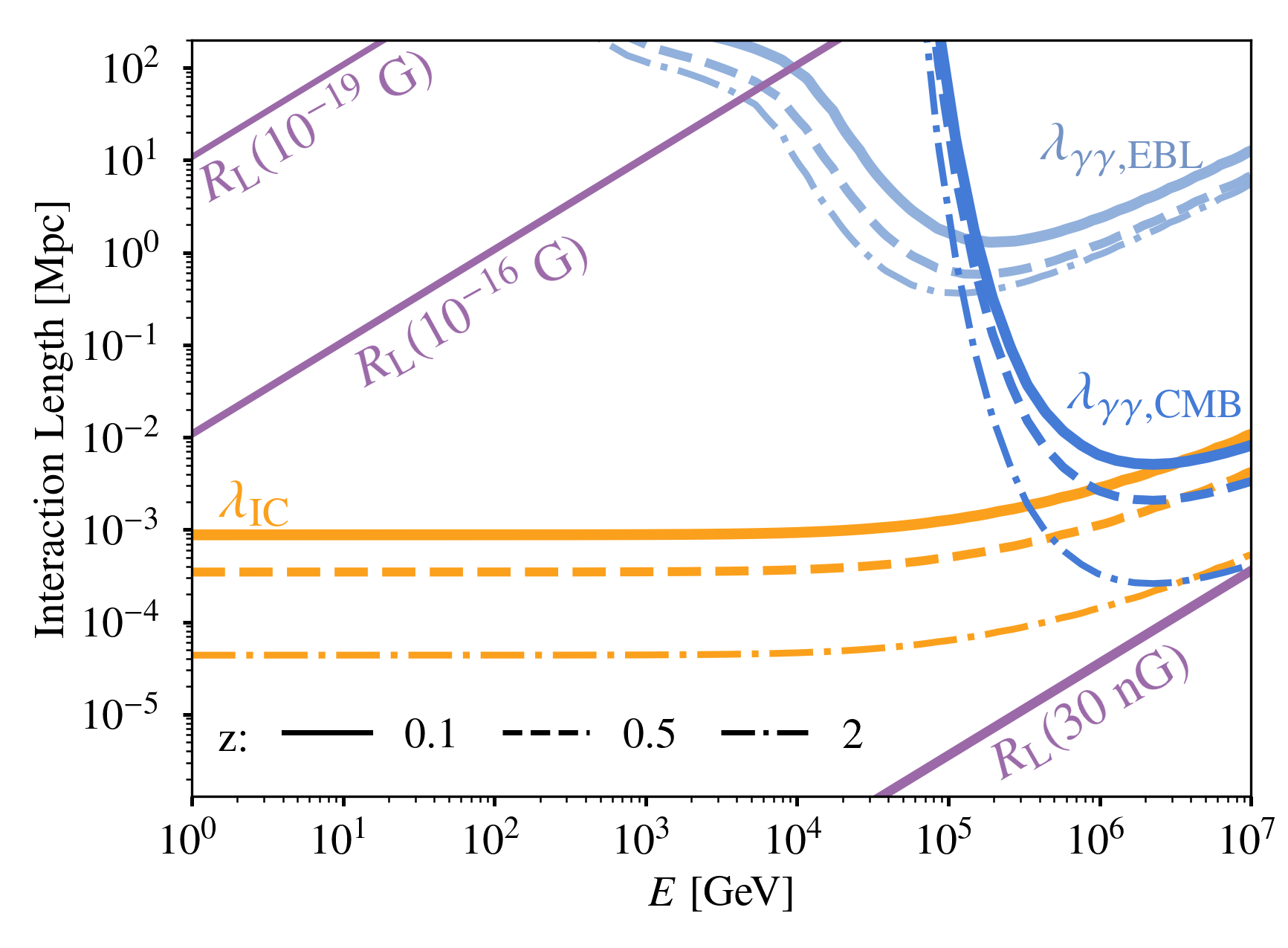}
    \caption{Energy dependence of key interaction lengths and Larmor radii relevant to the IGMF. The IC interaction length (\(\lambda_{\mathrm{IC}}\), orange) is nearly energy-independent in the Thomson regime, while the pair production lengths with EBL (\(\lambda_{\gamma\gamma,\mathrm{EBL}}\), light blue) and CMB (\(\lambda_{\gamma\gamma,\mathrm{CMB}}\), dark blue) show strong energy and redshift dependence, with curves shown for \(z = 0.1\) (solid), \(0.5\) (dashed), and \(2\) (dot-dashed). The Larmor radii \(R_L\) of electrons are shown in purple for magnetic field strengths representative of different environments: \(B_{\text{fil}}=30\,\mathrm{nG}\) for filaments, and \(B_{\text{void}} = 10^{-16}\,\mathrm{G}\) and \(10^{-19}\,\mathrm{G}\) for voids. Note that in our cases synchrotron energy losses are not important (see Refs.~\cite{Murase:2011yw,Sherman:2025gir}) for the importance of synchrotron cascades.}
    \label{fig: Interaction_length}
\end{figure}

In this study, we consider the sources with redshift $z\leq2$. Thus, the evolution of the background photon fields with redshift needs to be accurately accounted for. The EBL model used in this work is based on the empirical approach developed by Domínguez et al.~\cite{Dominguez:2010wxv}. This model uses observational data from the All-wavelength Extended Groth Strip International Survey (AEGIS), combined with galaxy spectral energy distribution types and the evolving rest-frame K-band galaxy luminosity function up to $z=4$. By considering the contributions from different galaxy types—quiescent, star-forming, starburst, and AGN—this model provides a robust estimate of the EBL over a wide range of wavelengths (0.1-1000 \unit{\mu m}).  

In addition to the EBL, the CMB provides an all-pervasive photon field originating at recombination ($z\simeq1100$). At redshift $z$ its temperature simply scales as $T(z) = T_{0}\,(1+z)\ $, with $T_{0} = 2.7255\,$K today. Its spectral intensity then follows the usual blackbody distribution at that temperature, peaking near millimeter wavelengths in the observer’s frame. 

\subsection{Magnetic Field Configuration}
Theoretical studies have long predicted the presence of IGMFs, and several mechanisms have been proposed for their formation~\cite{Durrer:2013pga}. 
Seed magnetic fields in the early Universe can be generated by processes related to inflation, cosmological phase transitions, and plasma physics~\cite{Turner:1987bw, Subramanian:2015lua, Subramanian:2019jyd}, while subsequent astrophysical processes can amplify these fields to the strengths observed today, such as by massive primordial halos~\cite{Diaz:2024poo}, first stars~\cite{Sur:2010xa, Xu:2008yb}, or galaxy formation~\cite{Kulsrud:1996km, Donnert:2008sn}. 
Since the amplification process is more frequent in filaments, both theories and observations support the existence of a stronger magnetic field in filaments compared to voids. Observations stacking low-frequency radio and X-ray maps suggest that filament magnetic fields are on the order of 30–60 nG~\cite{Vernstrom:2021hru}. 
Similarly, measurements of the dispersion in residual Faraday rotation indicate a field strength of about 32 nG~\cite{Carretti:2022tbj}. Statistical studies of Faraday rotation suggest that magnetic fields in filaments have a coherence length of a few hundred kpc~\cite{Akahori:2010ym}. Therefore, in this study, we consider a turbulent magnetic field with a Kolmogorov power spectrum and root-mean-square strength $B_{\text{fil}}$ = 30 nG with coherence length $l^{\text{fil}}_c$ = 100 kpc. The radius of a filament is approximately a few Mpc~\cite{Akamatsu:2017sng, Santiago-Bautista:2020gbf}. In this work, we assume a line-of-sight distance of 2 Mpc for the filament structure, corresponding to the distance from the source to the filament’s edge (see also Refs.~\cite{Murase:2011cy,Takami:2011nn}).  

Without amplification by plasma motion or compression, the magnetic field in cosmic voids remains much weaker than in structured regions. Being extremely weak and diffuse, it cannot be directly measured and is instead constrained indirectly through astrophysical effects such as $\gamma$-ray propagation and Faraday rotation~\cite{Durrer:2013pga, AlvesBatista:2021sln}. 
Observations based on delayed GeV $\gamma$-ray pair echoes set a lower bound of $10^{-19}$\,G on the IGMF~\cite{Vovk:2023qfk}, while others suggest a characteristic strength of approximately $10^{-16}$\,G~\cite{Neronov:2010gir, AlvesBatista:2021sln, Subramanian:2015lua}.

In this study, we employ $B_{\text{void}}$ strengths of $10^{-16}$ G and $10^{-19}$ G, with the same Kolmogorov power spectrum as mentioned earlier, to represent plausible scenarios for cosmic void environments. 
Finally, the coherence length is considered to be $l_c^{\text{void}}$ = 1 Mpc in our simulation.

\subsection{Particle Propagation and Interaction}
\begin{figure*}[t]
    \centering
    \includegraphics[width=0.95\textwidth]{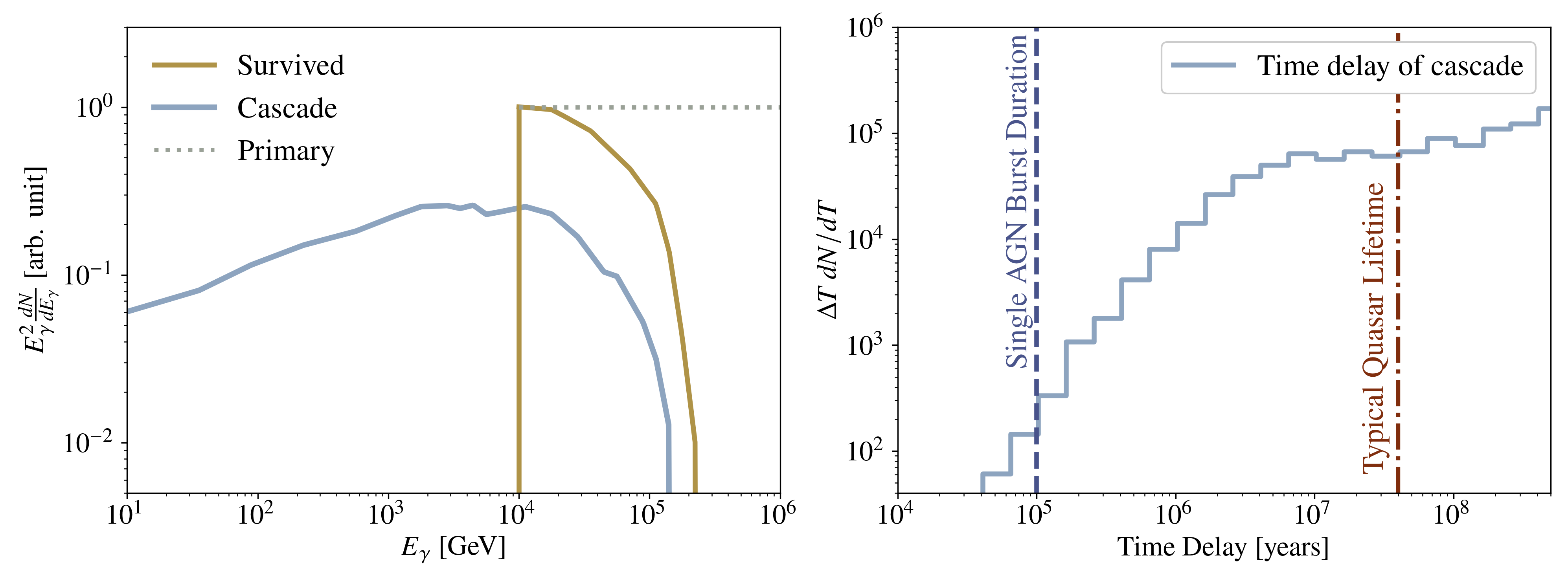}
    \caption{$\gamma$-ray propagation in a cosmic filament with \(B = 30\,\mathrm{nG}\) and radius = 2 Mpc at $z=0.1$. The left panel shows energy spectra of the $\gamma$-ray components, plotted in \(E^2\,dN/dE\). The dotted line shows the injected primary spectrum from 10 TeV to 1 PeV, the dark yellow solid line indicates the surviving primary $\gamma$-rays after propagation, and the blue solid line represents the cascade component produced via IC scattering. The right panel shows the time delay distribution of the cascade photons, indicating that secondary $\gamma$-rays arrive with delays ranging from \(10^4\) to over \(10^8\) years. As a comparison, the dashed dark blue line indicates a typical time of one AGN burst phase~\cite{Schawinski:2015xka}; the dot-dashed red line is the typical lifetime of luminous quasars~\cite{Martini:2000mj}.}
    \label{fig: filament_test}
\end{figure*}

During $\gamma$-ray propagation, we consider the primary physical processes that shape the development of electromagnetic cascades in the intergalactic medium: pair production ($\gamma \gamma \rightarrow e^+ e^-$) and IC scattering ($e^\pm \gamma_{\mathrm{CMB/EBL}} \rightarrow e^\pm \gamma$), along with magnetic deflection by the IGMF. 
A high-energy $\gamma$-ray emitted at angle $\theta$ relative to the observer's line of sight and traveling a luminosity distance $D_s$ can interact with low-energy background photons, initiating a cascade. In the end, photons are detected with a detection angle $\varphi$. This process and those relative parameters are demonstrated in Figure~\ref{fig: geo}. 

In the cosmological comoving frame, where the background photon field is isotropic, the proper interaction length of $\gamma$-ray photons or electrons in the CMB and EBL can be expressed by
\begin{equation}
\lambda^{-1}(E) = \int_{-1}^{1} \frac{1 - \mu}{2} d\mu \int_0^\infty d\varepsilon \, \frac{dn}{d\varepsilon} \, \sigma(E, \varepsilon, \mu),
\end{equation}
where $\sigma(E, \varepsilon, \mu)$ is the differential cross section for the interaction and $dn/d\varepsilon$ is the proper number density of target photons of energy $\varepsilon$. 
The outer integral represents an average over the interaction angle $\mu = \cos\psi$ under the assumption that the background photon field is isotropic. 
This is the average distance a particle can travel before interaction. By assuming the background photon fields are isotropic, the outer integral averages over the interaction angle $\mu$. 
Figure~\ref{fig: Interaction_length} shows the interaction length of each interaction at different redshifts. For pair production, the relevant cross section rises sharply above the threshold $E_\gamma\varepsilon(1-\mu)\gtrsim 2m_e^2c^4$, and the corresponding interaction length $\lambda_{\gamma\gamma}$ decreases rapidly with energy. At TeV energies, pair production is primarily governed by interactions with the EBL. Near $100~\mathrm{TeV}$, absorption by the far-infrared component of the EBL remains important, while interactions with the CMB become increasingly relevant at higher energies. The relative contributions of the two background photon fields depend on the $gamma$-ray energy and redshift, as illustrated in Figure~\ref{fig: Interaction_length}.

For electrons and positrons undergoing IC scattering, the interaction length, $\lambda_{\mathrm{IC}}$, is nearly energy-independent in the Thomson regime, due to the flat energy dependence of the cross section. Beyond $\sim$100 TeV, the IC process enters the Klein–Nishina regime, where the cross section decreases with energy~\cite{Jones:1965IC}. The electrons and positrons are deflected by the IGMF, with the deflection governed by the Larmor radius $R_L = \varepsilon_e/Be$.
The $R_L$ for $B_{\text{fil}}$ = 30 nG in filaments and $B_{\text{void}}$ = $10^{-16}$ or $10^{-19}$ G in voids are visualized in Figure~\ref{fig: Interaction_length}.

\subsection{Attenuation inside Filament}
After leaving the source, $\gamma$-rays are expected to propagate approximately 2 Mpc within the filament before entering the cosmic void. Photons with energy approaching and exceeding 100 TeV are expected to be strongly attenuated by the CMB and EBL photon fields, as their pair-production interaction length is shorter than the filament radius~\cite{Murase:2011yw}. For photons between 10 and 100 TeV, the level of attenuation depends strongly on the source redshift, since the EBL intensity increases with redshift up to 2, as shown in Figure~\ref{fig: Interaction_length}.
Because the coherence length $l_c^{\text{fil}}$ exceeds the IC interaction length $\lambda_{\text{IC}}$, electrons propagate through a nearly uniform magnetic field during their energy-loss phase, following smooth, deterministic arcs rather than experiencing random-walk deflections~\cite{Murase:2011yw}.
The fact that $R_L$ in the filament is much shorter than the IC interaction length, $\lambda_{\text{IC}}$, leads to large angular deflections of the electrons. As a result, the cascade radiation from the filament has a broad angular distribution and a huge time delay, making it difficult to associate with the original source. 

We have carried out a more detailed analysis via simulations. Figure~\ref{fig: filament_test} shows the results for a source located at redshift $z = 0.1$, where the primary $\gamma$-ray spectrum follows an $E^{-2}$ power law, extending from 10~TeV to 1 PeV. After propagating 2 Mpc through a filament and reaching a screen aligned with the line of sight, the resulting spectrum is shown in the left panel, which includes both the surviving primary photons and the secondary cascade component. As expected from the length-scale analysis, a spectral cutoff appears around $100~\mathrm{TeV}$ due to pair-production attenuation by the combined EBL and CMB photon fields. The right panel presents the time delay distribution of the cascade photons. Two vertical lines indicate characteristic AGN timescales: the duration of a single accretion burst~\cite{Schawinski:2015xka} and the typical lifetime of luminous quasars~\cite{Martini:2000mj}. The majority of cascade photons arrive with delays exceeding $10^5$ years, comparable to or even longer than these AGN lifetimes. As a result, the cascade emission would arrive long after the central engine has faded, making it unassociated with either transient or steady sources. Therefore, only the $\gamma$-ray component that successfully escapes from the structured region is considered detectable by $\gamma$-ray telescopes in this scenario. 

\section{Multimessenger Interface and the Origin of IceCube Neutrinos}
\label{sec:MMinterface}

\subsection{IceCube Detectability Criterion}

\begin{figure*}[t]
    \centering
    \includegraphics[width=0.95\textwidth]{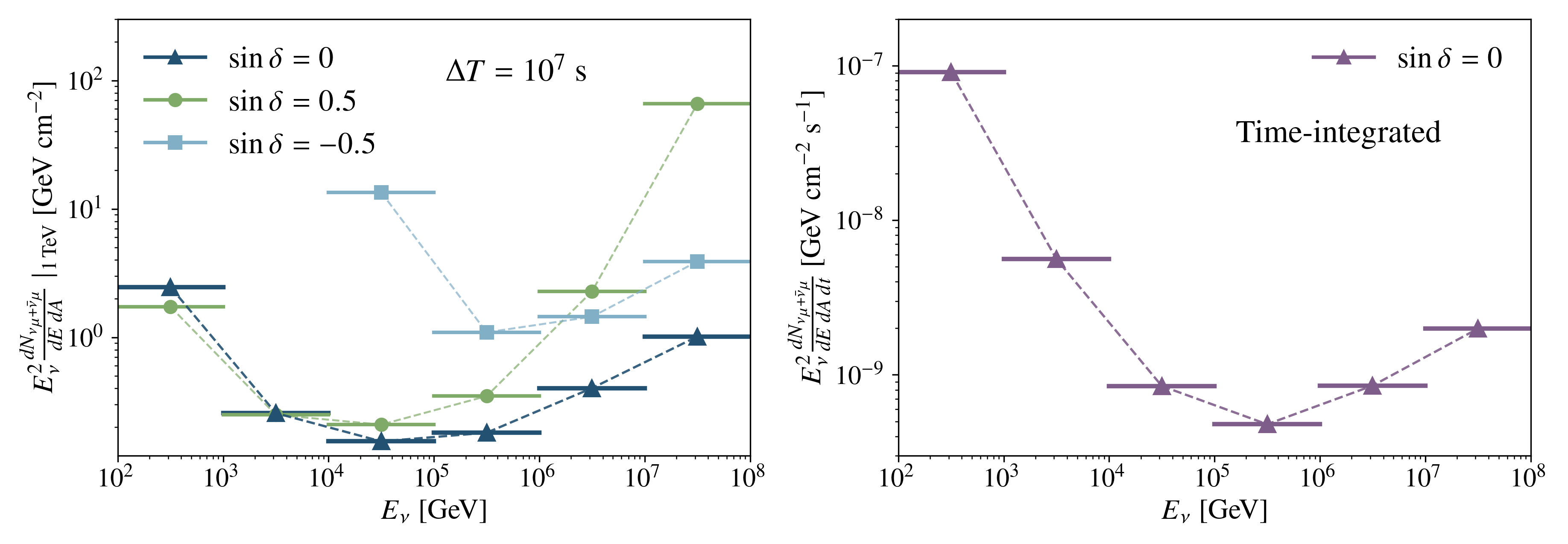}
    \caption{
    IceCube point-source sensitivity for both transient (left) and steady emission (right) cases. Left panel shows time-integrated sensitivity per energy bin at different zenith angles, assuming an $E^{-2}$ spectrum within each bin, and a fixed flare duration of $\Delta T = 10^7$~s~\cite{IceCube:2020mzw}. The right panel shows the differential sensitivity at the horizon, assuming an unbroken $E^{-2}$ power-law over the entire energy range, used for steady emission~\cite{Pizzuto:2022ijm}. 
    }
    \label{fig:IC_sensitivity}
\end{figure*}

To evaluate the detectability of a $gamma$-ray counterpart of high-energy cosmic neutrinos, we calculate the expected $\gamma$-ray flux associated with a neutrino counterpart in specific IceCube, as the primary operating neutrino telescope, detection scenarios.
We consider two distinct regimes: transient and steady emissions. For transients, inspired by the 2014--2015 neutrino flare from TXS~0506+056, we adopt a time window of $\Delta T = 10^7$ s, approximately 115 days.
For transients, inspired by the 2014--2015 neutrino flare from TXS 0506+056, we adopt a time window of $\Delta T = 10^7$ s, approximately 115 days. Although IceCube's time-integrated sensitivity can remain approximately constant for shorter time windows in the low-background regime, the detectability of gamma-ray counterparts also depends on the propagation-induced time delay of cascade photons. A shorter time window would impose a more restrictive time-delay selection, potentially suppressing the observable cascade emission. Therefore, we restrict our transient analysis to the benchmark $\Delta T = 10^7$ s.
In contrast, for steady emission, we adopt an idealized
stationary-source scenario in which the emission remains
approximately constant over timescales sufficiently long
for the relevant electromagnetic cascades to reach a
stationary state. Under this assumption, propagation-induced
time delays do not suppress the time-averaged cascade
flux, since delayed photons are continuously replenished
by earlier emission. Therefore, no explicit time-delay
cut is imposed, although the emission angle $\theta$
and detection angle $\phi$ are still constrained, as
discussed in Section~\ref{sec:Method}.

IceCube is primarily sensitive to neutrinos in the TeV to PeV energy range, though the specific energies that contribute most to a source search depend on the event selection criteria, the source's location in the sky, and the spectral shape of the signal. 
The IceCube muon-neutrino dataset achieves its best sensitivity near the celestial horizon because this region offers an optimal balance
of signal over background ratio
without significant Earth absorption of neutrinos, which allows efficient detection with minimal background interference. In this work, we use IceCube's sensitivity from the horizon as a benchmark. 

The right panel of Figure~\ref{fig:IC_sensitivity} shows such differential sensitivity at the horizon for a time-integrated point source analysis~\cite{Pizzuto:2022ijm}. For transients, the differential sensitivity with a $\Delta T=10^7$~s time window will be used. By assuming a constant emission in time window $\Delta T$, the fluence can be expressed as $dN/dEdA=(dN/dEdAdt)\times\Delta T$. The left panel of Figure~\ref{fig:IC_sensitivity} shows the time-integrated sensitivity with $\Delta T=10^7$~s at different zenith angles, and, as we just discussed, the best sensitivity happens at the horizon~\cite{IceCube:2020mzw}. 

\subsection{Pionic $\gamma$-rays}
 For the given neutrino flux at the level of IceCube’s point-source sensitivity in a specific energy bin, we estimate the corresponding  $\gamma$-ray flux at the source under the assumption of hadronic production. To begin, we convert the observed Earth-frame neutrino flux into the source-frame differential production rate $Q$. Let \( E_\nu\) be the neutrino energy at Earth, and \( \varepsilon_\nu = (1+z) E_\nu \) the neutrino energy at the source. 

 The source-frame neutrino energy flux (per unit time) is then given by:

\begin{equation}
\left. \varepsilon_\nu {\mathcal E}_{\varepsilon_{\nu_\alpha}}(\varepsilon_\nu) \right|
= 4\pi D_p^2(z) (1+z) \cdot \left. E_\nu^2 \phi_{\nu} \right|,
\label{eq:redshift_conversion}
\end{equation}
where $D_p(z)$ is source's proper distance and $E_\nu^2 \phi_{\nu}$ is the energy fluence.
Note that in this expression, the left-hand side is evaluated at \(\varepsilon_\nu\) and the right-hand side at \(E_\nu\). In an \(E^2 dN/dE dt\) representation, this transformation shifts the spectrum horizontally by a factor of \((1+z)\) in energy and vertically by the same factor due to the redshift energy losses
\begin{equation}
E_\gamma^2 \int dt \frac{dN_\gamma}{dE_\gamma dt} = E_\gamma^2 \phi_{\gamma},
\end{equation}
where the time integral runs over the time-delay.

High-energy cosmic neutrinos and $\gamma$-rays are jointly produced through charge and neutral pion decay from hadronic interactions. 
Two primary channels contribute: proton-proton ($pp$) and proton-photon ($p\gamma$) collisions. In pp interactions, comparable numbers of charged and neutral pions are produced, leading to similar total energy in neutrinos and $\gamma$-rays. In contrast, \(p\gamma\) interactions proceed predominantly through the \(\Delta^+\) resonance, which decays preferentially into a neutron and a charged pion, while the alternative decay into a proton and a neutral pion is subdominant. 

In this work, we focus on the \(p\gamma\) scenario. This choice is motivated by the physical conditions of many proposed neutrino sources, such as blazars, $\gamma$-ray bursts, and tidal disruption events, which are characterized by intense photon fields and low baryon densities. This choice will offer a conservative neutrino flux expectation compared to the hadronuclear assumption ($pp$) channel.
Assuming a transparent source environment, the source-frame $\gamma$-ray production rate ($Q$) can be estimated from the neutrino production rate using the relation~\cite{Murase:2015xka}:
\begin{equation}
\frac{1}{3} \sum_{\alpha} \varepsilon_\nu {\mathcal E}_{\varepsilon_{\nu_\alpha}} \approx \frac{K_\pi}{4} \cdot [\varepsilon_\gamma {\mathcal E}_{\varepsilon_\gamma} ]_{\varepsilon_\gamma=2\varepsilon_\nu}
\label{eq:nugamma_relation}
\end{equation}
Here, \(K_\pi\) is the charged-to-neutral pion ratio, with \(K_\pi \approx 1\) for \(p\gamma\) interactions. The factor \(1/4\) accounts for the approximate relation \(\varepsilon_\gamma \approx 2 \varepsilon_\nu\) and energy divisions in pion decay.

Starting from IceCube sensitivity, we calculate the neutrino flux at production based on Equation~\ref{eq:redshift_conversion}; then, under the assumption of $\gamma$-ray transparent source, the $\gamma$-ray flux from the source can be evaluated by Equation~\ref{eq:nugamma_relation}, and the $\gamma$-ray flux reaching Earth can be calculated from simulation introduced in Section~\ref{sec: Simulation_intro}.
\section{Method}
\label{sec:Method}
\begin{figure*}[t]
    \centering
    \includegraphics[width=0.94\textwidth]{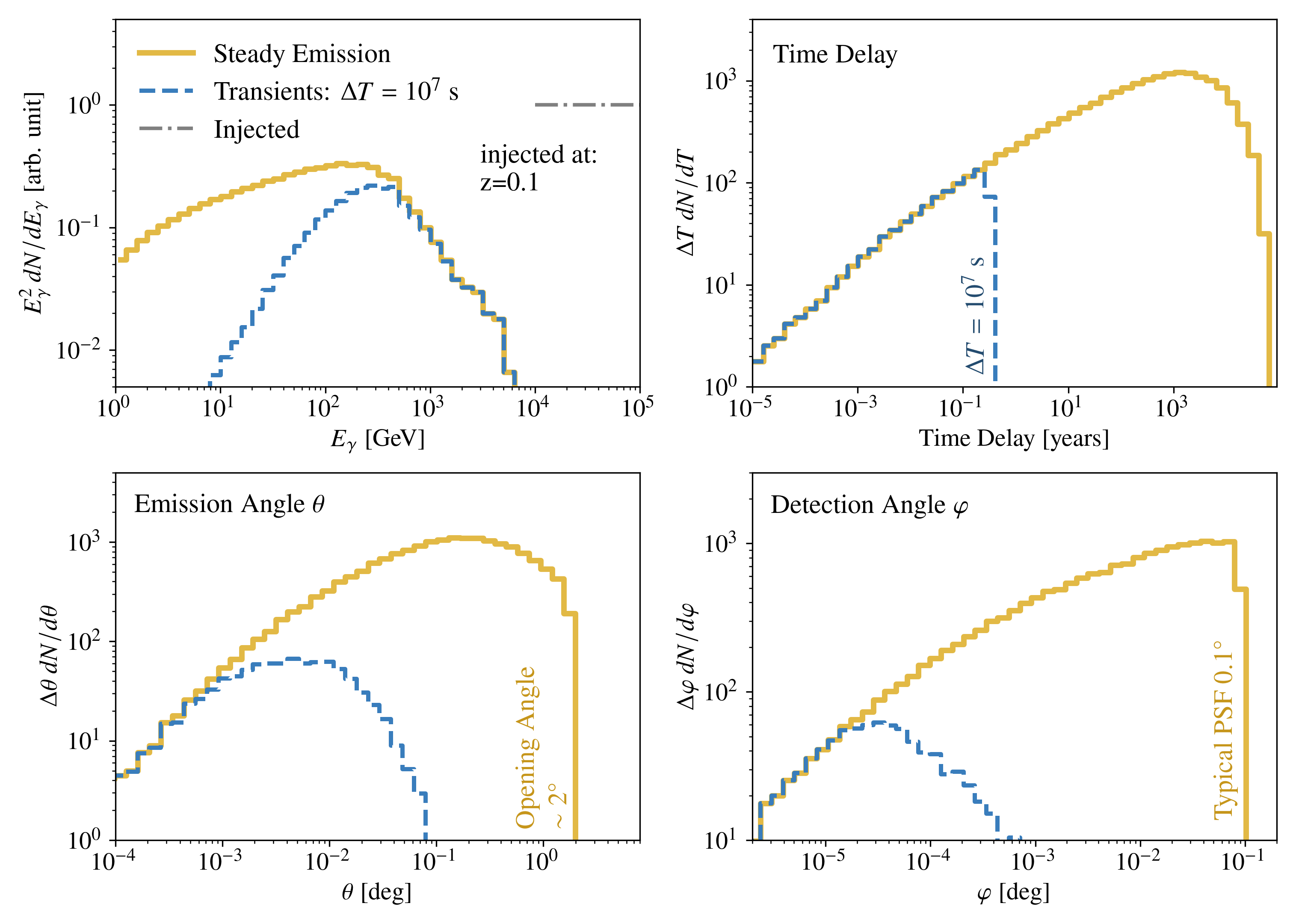}
    \caption{
    Cascade photon distributions from a primary source located at redshift \(z = 0.1\) with an $\varepsilon_\gamma^{-2}$ spectrum from 10 TeV to 100 TeV. Top left: Energy spectrum of the cascade photons. The injected spectrum (gray dash-dotted) produces the unfiltered cascade (yellow solid) and the subset consistent with transient emission over \(\Delta T = 10^7\,\mathrm{s}\) (blue dashed). Top right: Time delay distribution, showing that most cascade photons arrive with delays exceeding the transient window \(\Delta T = 10^7\,\mathrm{s}\), especially beyond \(\sim 10^2\,\mathrm{years}\). Bottom left: Emission angle \(\theta\) relative to the line of sight. We set a $2^\circ$ constraint based on the typical jet opening angle. Bottom right: Detection angle \(\varphi\), showing the distribution at Earth. We constrain $\varphi$ to be smaller than $0.1^\circ$, which is approximately \textit{Fermi}-LAT's PSF above GeV~\cite{Fermi-LAT:2013mql}. The energy spectrum on the top left is normalized with the injected component set to unity, while the other three panels are not normalized. 
    }
    \label{fig:method01_z01}
\end{figure*}

\subsection{Geometric Analysis}

\begin{figure*}[t]
    \centering
    \includegraphics[width=0.94\textwidth]{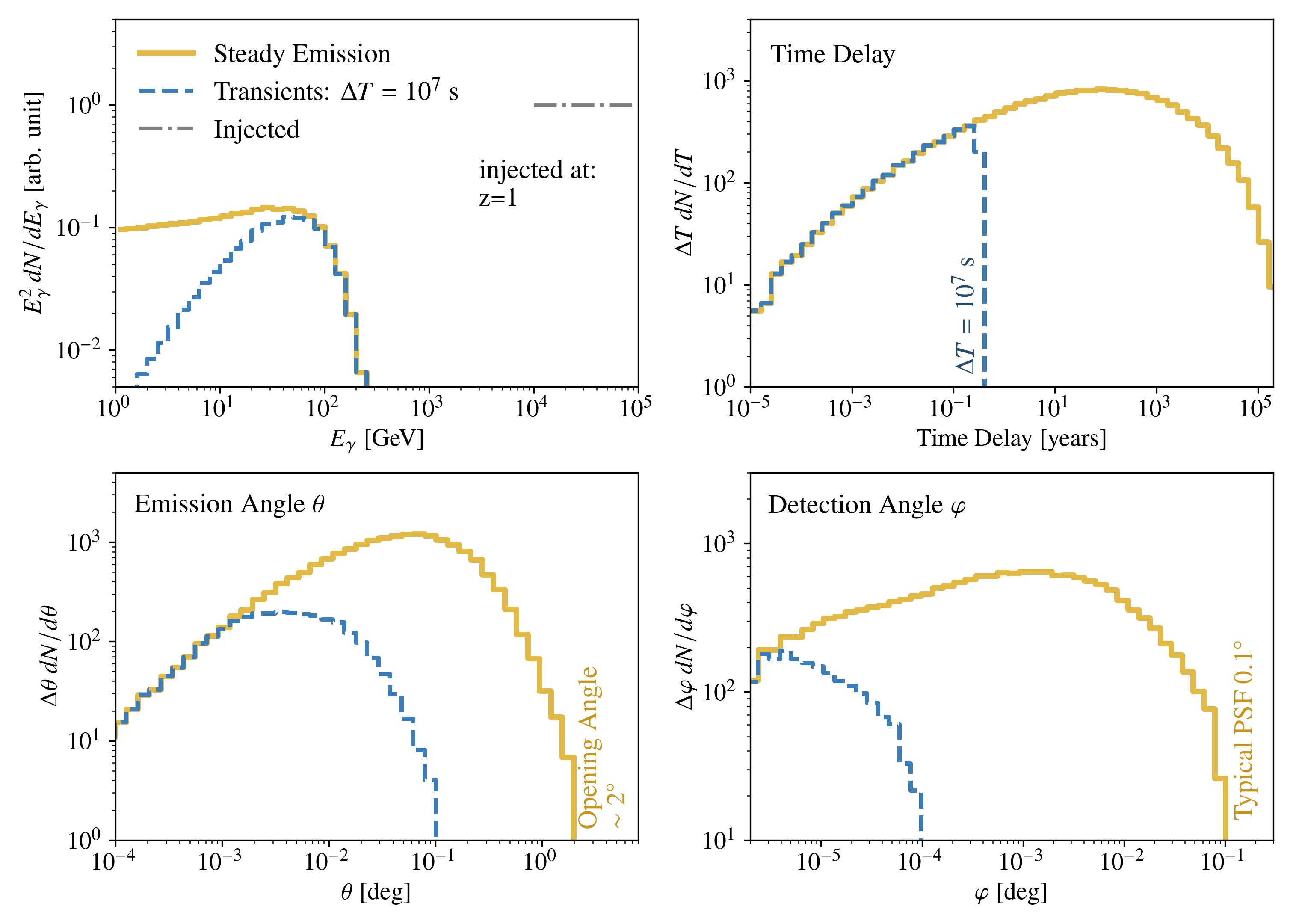}
    \caption{
    Same as Figure \ref{fig:method01_z01}, but for a source located at redshift \(z = 1\).
    }
    \label{fig:method01}
\end{figure*}
The detectability of $\gamma$-rays is strongly influenced by geometric effects, including halo formation and time delay. Figure~\ref{fig: geo} illustrates the geometry of cascade development as VHE $\gamma$-rays propagate.
For cosmological propagation, multi-generation cascades can develop. The emission angle  $\theta$ is defined as the angle between the line of sight and the initial direction of the primary $\gamma$-ray whose cascade leads to a photon that reaches Earth. The detection angle $\varphi$ is the angle between the line of sight and the arrival direction of that photon at Earth.

Due to attenuation in the filament, the highest-energy $\gamma$-rays that reach the void are limited to around 100~TeV. Photons above $\sim$ 10~TeV interact with the EBL and quickly initiate first-generation cascades. Since the IC photon energy scales with the square of the electron energy, and the electrons follow the primary $\gamma$-ray spectrum, this quadratic relation compresses the resulting spectrum~\cite{Fitoussi:2017ble}. As a result, the photon spectrum becomes harder than the injection and follows approximately $dN/dE \propto E^{-1.5}$.

In the Thomson regime, the typical photon energy in the first-generation cascade is given by
\begin{align}\nonumber
    \varepsilon_\gamma^{\text{1st}} &= \frac{4}{3} 
    \left( \frac{\varepsilon_e}{m_e c^2} \right)^2 
    \varepsilon_{\text{CMB}} \\
    &\approx 8.9 \times 10^{-7}\mathrm{GeV} \cdot (1+z) \cdot 
    \left( \frac{\varepsilon_\gamma^0}{\text{1 GeV}} \right)^2,
\end{align}

where $\varepsilon_e$ is electron's energy assumed to be $\varepsilon_\gamma^0/2$, $\varepsilon_{\text{CMB}}$ is the typical CMB photon energy in the cosmic rest frame, and $\varepsilon_\gamma^0$ is the energy of the primary photon.

To produce cascade photons with at least GeV energies, the electron must have energies above $\sim$ 1 TeV.
The relevant deflection length scale is the cooling distance $D_{\text{IC}}$, defined as the characteristic distance over which an electron loses a significant fraction of its energy via IC scattering off CMB photons. It can be expressed by
\begin{equation}
    D_{\text{IC}}=\frac{3 m_e^2 c^4}{4 \sigma_T u_{\mathrm{CMB}} \varepsilon_e},
\end{equation}
where $\sigma_T$ is the Thomson cross section and $u_{\mathrm{CMB}}$ is the CMB's energy density, which depends on redshift (see Figure~\ref{fig: Interaction_length}). 
We typically have $R_L\ggg D_{\rm IC}$. Thus, the deflection angle $\delta_{\text{1st}} \approx D_{\text{IC}} / R_L$, yielding $\delta_{\text{1st}} \sim {5\times10^{-3}}^\circ$ for 20~TeV electrons in a void field of $B_{\text{void}} = 10^{-16}$\,G, and $\delta_{\text{1st}} \sim {10^{-5}}^\circ$ for $B_{\text{void}} = 10^{-19}$\,G. In a single-generation cascade, the total deflection angle $\delta_{\text{1st}}$ is the sum of the emission angle $\theta$ and detection angle $\varphi$.

In propagation over hundreds of Mpc, the first-generation cascade becomes attenuated, and the second-generation cascade gradually emerges, producing photons with energies around hundreds of GeV and lower. In this energy range, \( \lambda_{\gamma\gamma,\text{EBL}} \) exceeds \( 10^4 \) Mpc, which is much larger than the redshift distances considered in this study. As a result, the third-generation cascade contributes only marginally to the observed spectrum at a few GeV. The electron–positron pairs involved in IC scattering are typically at the TeV scale. For such electrons, the characteristic deflection angles are \( \delta_{\text{2nd}} \sim 0.5^\circ \) for \( B_{\text{void}} = 10^{-16} \, \mathrm{G} \) and \( \delta_{\text{2nd}} \sim 5 \times {10^{-4}}^\circ \) for \( B_{\text{void}} = 10^{-19} \, \mathrm{G} \). When the source distance is much larger than $\lambda_{\gamma\gamma, \mathrm{EBL}}$, the cascades typically initiate near the source rather than near Earth. Therefore, the emission angle $\theta \sim \delta_{\text{2nd}}$ and the detection angle $\varphi\ll \delta_{\text{2nd}}$.

When considering sources such as blazars, the opening angle of the \(\gamma\)-ray jet is typically a few degrees~\cite{Finke:2018pkl}. In this work, we adopt \( 2^\circ \) as an upper limit on the emission angle \( \theta \). The point-spread function (PSF) of \textit{Fermi}-LAT above GeV energies is approximately \( 0.1^\circ \), which is used as the upper limit on the detection angle \( \varphi \)~\cite{Fermi-LAT:2013mql}. For \( B_{\text{void}} = 10^{-19} \, \mathrm{G} \), the deflection angles \( \delta_{\text{1st}} \) and \( \delta_{\text{2nd}} \) are much smaller than these limits. However, for \( B_{\text{void}} = 10^{-16} \, \mathrm{G} \), the deflection becomes comparable to the observational constraints and requires more careful treatment.
This is shown in Figure~\ref{fig:method01_z01} and Figure~\ref{fig:method01}, where the simulation results are presented for sources at redshifts $z=0.1$ and $z=1$, respectively. Here, we adopt a void magnetic field of \( B_{\text{void}} = 10^{-16} \, \mathrm{G} \). 
The yellow solid line represents the steady emission scenario, where constraints are applied to the emission angle $\theta$ and detection angle $\varphi$. The blue dashed line corresponds to the transient scenario with $\Delta T = 10^7$\,s, where a time-delay filter is applied. The gray dash-dotted line shows the injected spectrum from 10 to 100~TeV, which is almost completely attenuated at both redshifts.
At $z = 0.1$, the second-generation cascade starts to appear, contributing to the spectrum below 100\,GeV. 
In contrast, at $z = 1$, the second-generation cascade is fully developed and dominates the low-energy range. 
In the case of transients, applying the time-delay filter removes many low-energy photons. 
These photons are typically produced by lower-energy electrons, which are more easily deflected and arrive with longer delays. For transients, the time-delay filter supersedes the $\theta$ and $\varphi$ filters, effectively restricting the values of the latter two to be much smaller than the typical jet opening angle and the detector's PSF, respectively. 

In the steady emission scenario, the constraints on both $\theta$ and $\varphi$ become important, especially for sources at lower redshifts, such as $z = 0.1$ (see Figure~\ref{fig:method01_z01}). At this redshift, the large deflection of second-generation cascade electrons results in a large $\varphi$.
However, at higher redshifts, $\varphi$ tends to be smaller. This is because the higher density of CMB leads to a shorter IC cooling distance, following $D_{\text{IC}} \propto (1+z)^{-4}$, which limits the magnetic deflection of the electrons. 
Furthermore, at high redshifts, cascades are typically produced closer to the source than to Earth, which further compresses the observed detection angle $\varphi$. 
These effects explain why the angular distribution shown in the bottom panels of Figure~\ref{fig:method01_z01} is broader than that in Figure~\ref{fig:method01}.
\begin{figure*}[t]
    \centering
    \includegraphics[width=0.9\textwidth]{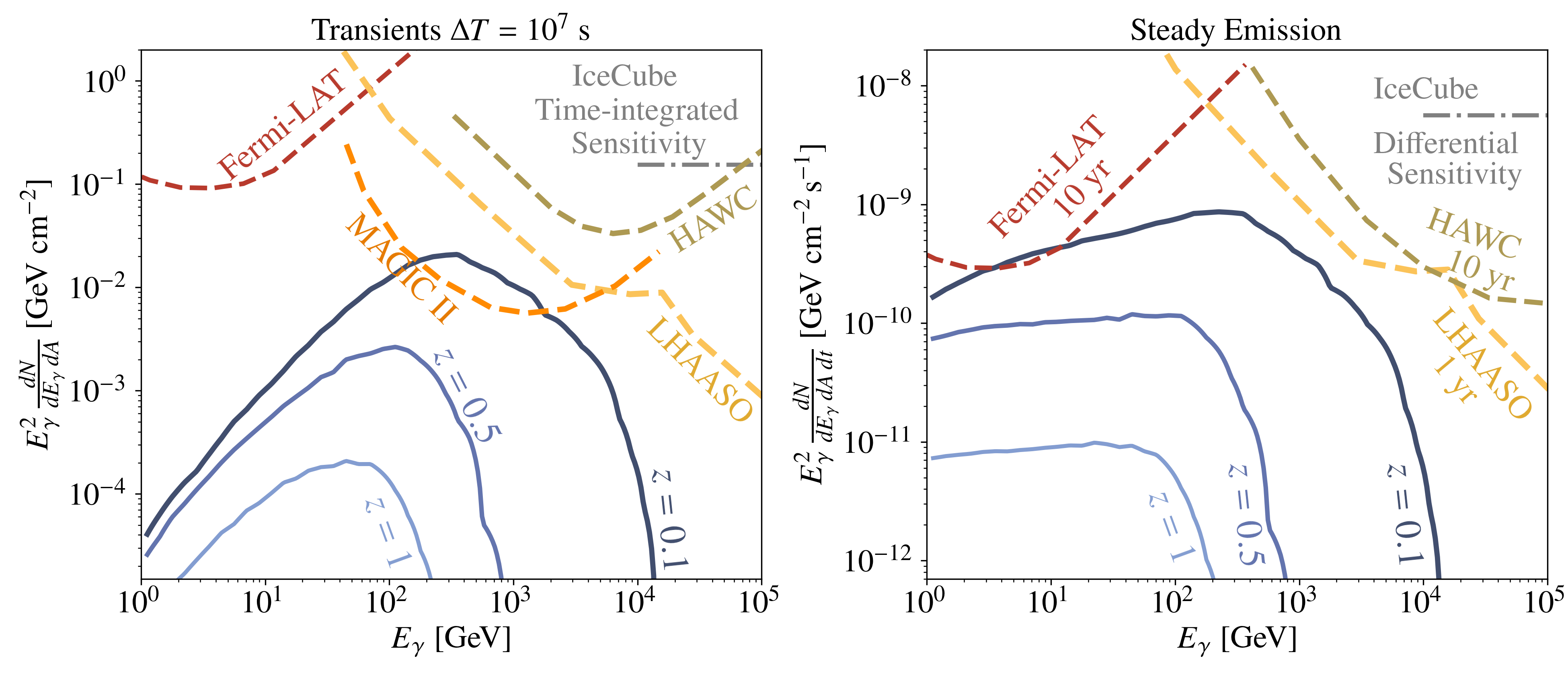}
    \caption{
    Comparison between the predicted cascade \(\gamma\)-ray fluence (left) or flux (right) associated with IceCube-detectable neutrino sources and the sensitivity thresholds of current \(\gamma\)-ray observatories. Left: transient scenario assuming a flare duration of \(\Delta T = 10^7\,\mathrm{s}\). Right: steady emission case. The solid curves represent the expected \(\gamma\)-ray fluence or flux at Earth for sources located at different redshifts (\(z = 0.1, 0.5, 1\)), derived under the assumption of hadronic co-production with neutrinos at the IceCube sensitivity limit. The dashed lines show the published point-source sensitivities for IceCube, \textit{Fermi}-LAT~\cite{Fermi-LAT:2012fsm, Funk:2013xds}, LHAASO~\cite{DiSciascio:2016rgi}, HAWC~\cite{Abeysekara:2017mjj, HAWC:2024plu}, and MAGIC II~\cite{MAGIC:2014zas}. This comparison reveals how $\gamma$-ray observability is affected by redshift and emission timescale, illustrating the challenge of detecting $\gamma$-ray counterparts to distant or short-lived neutrino sources.
    }
    \label{fig:method02}
\end{figure*}
\subsection{Gamma-Ray Counterpart to IceCube-Detectable Neutrinos}

Having established the criteria for cascaded
$\gamma$-rays in transient and steady emission
scenarios, we now assess whether the associated
$\gamma$-ray emission would be detectable by
current-generation $\gamma$-ray observatories.
To do so, we inject $\gamma$-rays at the source
assuming hadronic co-production with neutrinos
at the IceCube point-source sensitivity threshold,
as discussed in Section~\ref{sec:MMinterface}.

For transient emission, we express the observable
$\gamma$-ray signal in terms of fluence, defined
as the time-integrated flux over the relevant
observation window,
\begin{equation}
    \frac{dN_\gamma}{dE_\gamma\,dA}
    =
    \int_{\Delta T} dt\,
    \frac{dN_\gamma}{dE_\gamma\,dA\,dt}.
\end{equation}
In our simulations, we approximate the observable
transient cascade by selecting photons with
propagation-induced time delays smaller than
$\Delta T=10^7$~s. For steady emission, we instead
consider the differential $\gamma$-ray flux,
$dN_\gamma/dE_\gamma\,dA\,dt$, without imposing
an explicit time-delay cut.

The predicted cascade $\gamma$-ray fluence
or flux at Earth is then compared with the
sensitivity of instruments such as
\textit{Fermi}-LAT, HAWC, LHAASO, and MAGIC.
For these observatories, we adopt their
published point-source sensitivities as
reference benchmarks. These sensitivities do
not necessarily correspond to the detection
thresholds for the assumed transient duration
of $\Delta T=10^7$~s, since instrumental
sensitivity depends on the observation time
and background. Therefore, the comparison
provides a benchmark estimate of $\gamma$-ray
detectability rather than a duration-matched
transient sensitivity.

Figure~\ref{fig:method02} compares the expected $\gamma$-ray spectra for 3 different redshifts with these instruments' sensitivities.  
The dash-dotted gray line shows IceCube's differential sensitivity, which is also used to calculate the corresponding $\gamma$-ray emission at the source. 
The $\gamma$-rays then propagate through both the filament and the void, arriving at Earth with fluxes shown by the blue solid lines. In both the transient and steady cases, we present the resulting \(\gamma\)-ray spectrum for three representative redshifts: \(z = 0.1\), 0.5, and 1. 
As expected, the observable flux decreases with increasing redshift due to EBL attenuation within the filament. 
Additional attenuation in the void and adiabatic energy losses also contribute. 
In the transient scenario, a large portion of the cascade flux is removed by the time-delay cut, leading to strong suppression at low energies.
This significantly impacts detectability by \textit{Fermi}-LAT. In contrast, the steady emission scenario retains more of the cascade, producing a larger flux across different energies.
On the other hand, the detectability for ground-based $\gamma$-ray telescopes, which are sensitive to $\sim$ TeV energies and above, is primarily determined by the energy cut-off in the spectrum.

\section{Results}
\label{sec:Results}
\begin{figure*}[t]
    \centering
    \includegraphics[width=1\textwidth]{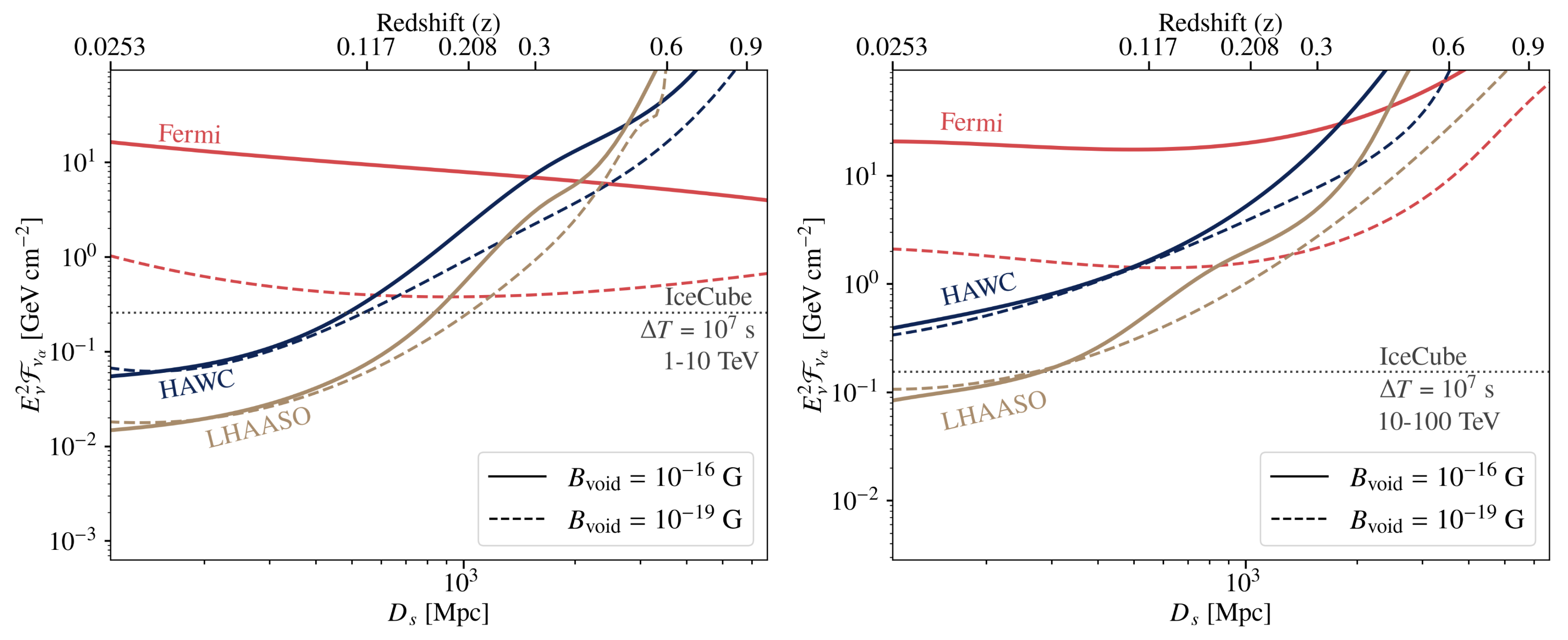}
    \caption{Required per-flavor neutrino fluence at Earth,
$E_\nu^2\mathcal{F}_{\nu_\alpha}$, as a function of source's luminosity distance $D_s$ (bottom axis) and redshift (top axis), for producing a $\gamma$-ray counterpart detectable by current $\gamma$-ray observatories in the transient scenario. The black dotted line shows IceCube’s point-source sensitivity in the corresponding energy bin. The left panel corresponds to the \( 1\text{--}10 \, \mathrm{TeV} \) energy bin, and the right panel to that of the \( 10\text{--}100 \, \mathrm{TeV} \) bin. Colored curves show the threshold neutrino fluence required for the resulting cascade $\gamma$-ray fluence to reach the point-source sensitivity of \textit{Fermi}-LAT (red), HAWC (blue), and LHAASO (brown) for 1 year of observation. Two IGMF strengths are considered: \( B_{\mathrm{IGMF}} = 10^{-16} \, \mathrm{G} \) (solid) and \( 10^{-19} \, \mathrm{G} \) (dashed). Regions above a $\gamma$-ray detector curve correspond to sources whose associated cascade $\gamma$-ray emission exceeds the adopted sensitivity benchmark, assuming $\gamma$-ray transparency at the source and hadronic co-production. Regions above both $\gamma$-ray detector curve and IceCube sensitivity indicates the multimessenger horizon.}
    \label{fig:T_binned}
\end{figure*}
\begin{figure*}[t]
    \centering
    \includegraphics[width=1\textwidth]{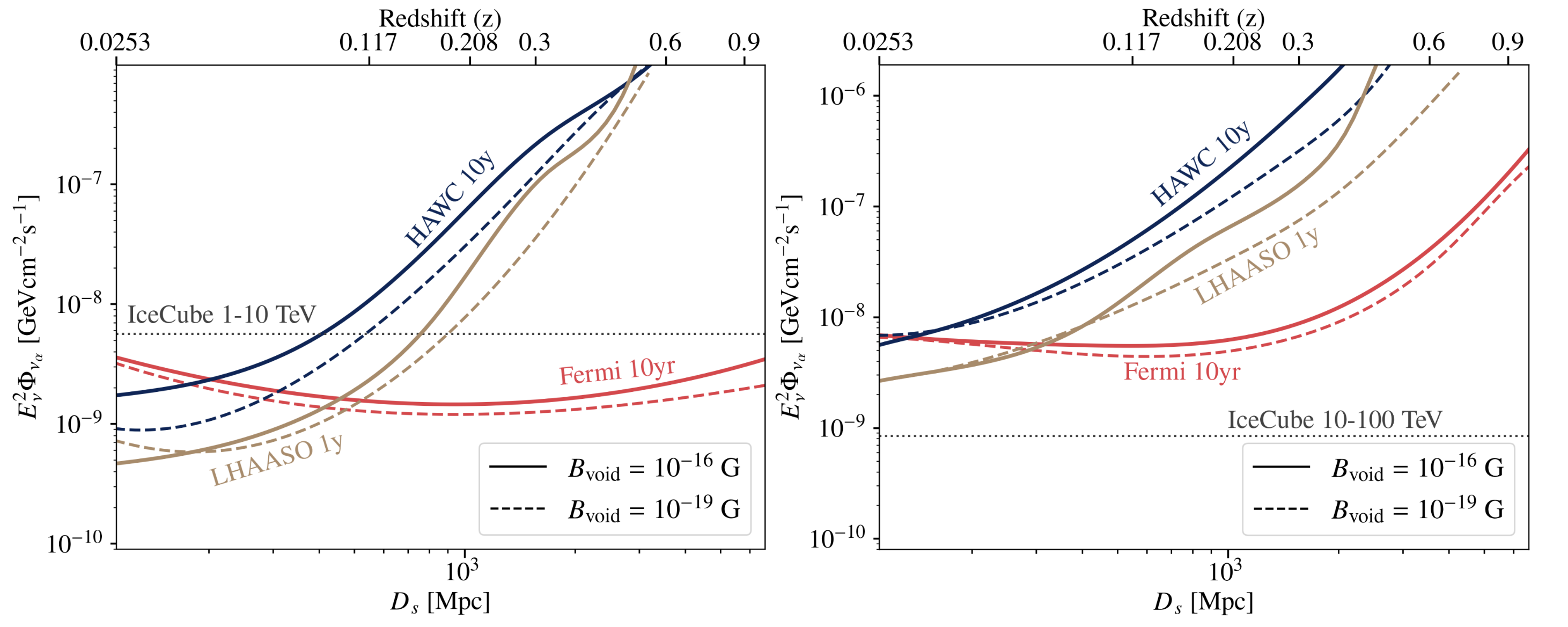}
    \caption{Required per-flavor neutrino flux at Earth, \( E^2 \Phi_{\nu_\alpha} \), as a function of the source's luminosity distance $D_s$ (bottom axis) and redshift (top axis), for producing a cascade $\gamma$-ray counterpart detectable by current $\gamma$-ray observatories under the steady emission scenario. The left panel corresponds to the \( 1\text{--}10 \, \mathrm{TeV} \) energy bin, and the right panel to \( 10\text{--}100 \, \mathrm{TeV} \). Colored curves represent the neutrino flux levels that would yield a $\gamma$-ray flux at the point-source sensitivity threshold of \textit{Fermi}-LAT (red, 10 years), HAWC (blue, 10 years), and LHAASO (brown, 1 year), assuming hadronic co-production and $\gamma$-ray transparency at the source. Solid and dashed curves correspond to intergalactic magnetic field strengths of \( B_{\mathrm{IGMF}} = 10^{-16} \, \mathrm{G} \) and \( 10^{-19} \, \mathrm{G} \), respectively. The black dotted horizontal lines indicate IceCube’s 10-year point-source sensitivity for each energy bin.
}
    \label{fig:S_binned}
\end{figure*}
\begin{figure*}[t]
    \centering
    \includegraphics[width=1\textwidth]{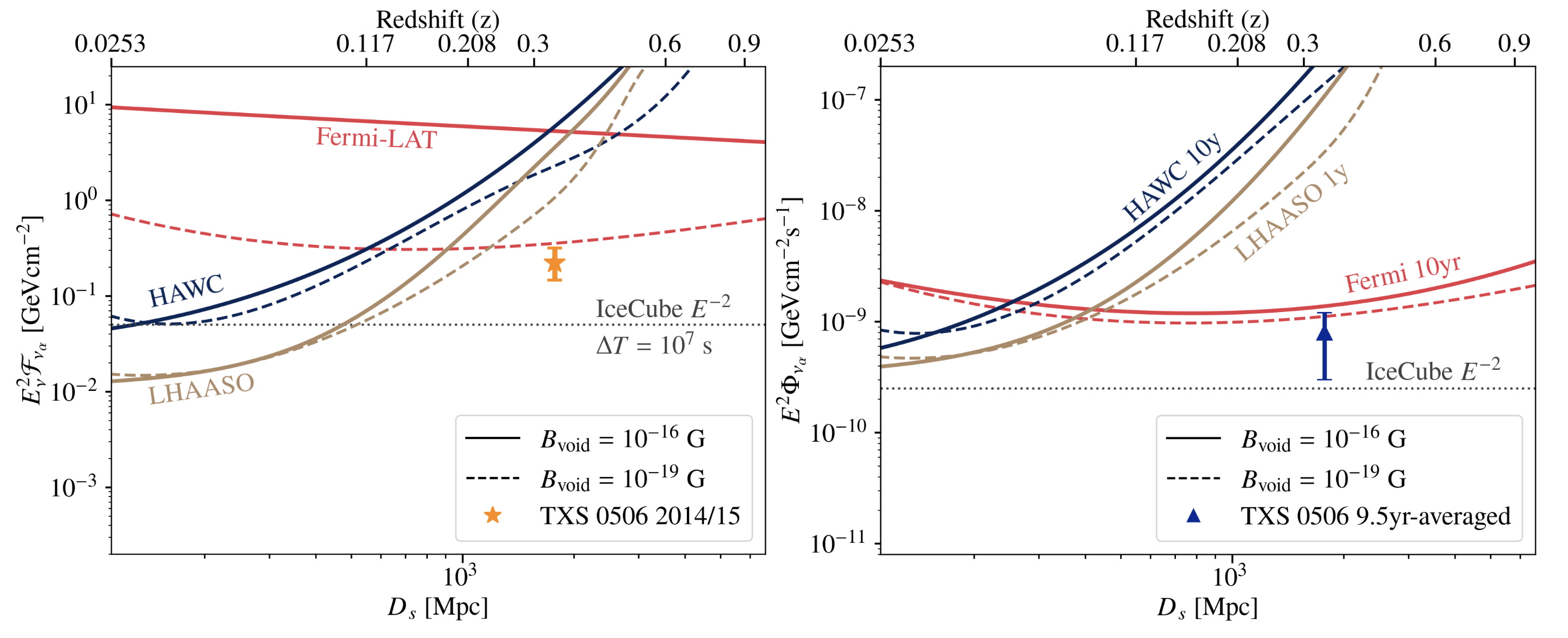}
    \caption{Required per-flavor neutrino fluence,
$E_\nu^2\mathcal{F}_{\nu_\alpha}$ (left), and flux,
$E_\nu^2\Phi_{\nu_\alpha}$ (right), at Earth as a function of source's luminosity distance $D_s$ (bottom axis) and redshift (top axis), for $\gamma$-ray cascade detection under the assumption of a continuous unbroken \( E^{-2} \) power-law neutrino spectrum. The left panel applies transient filtering (\( \Delta T = 10^7 \, \mathrm{s} \)), while the right panel corresponds to the steady case. Colored curves represent the minimum neutrino flux required for the cascade $\gamma$-ray flux to be detectable by \textit{Fermi}-LAT (red), HAWC (blue), and LHAASO (brown), assuming hadronic co-production and source transparency. Solid and dashed curves show results for intergalactic magnetic fields \( B_{\mathrm{IGMF}} = 10^{-16} \, \mathrm{G} \) and \( 10^{-19} \, \mathrm{G} \), respectively. The horizontal black dotted lines indicate IceCube’s point-source sensitivity assuming an unbroken \( E^{-2} \) spectrum. Benchmarks for the blazar TXS 0506+056 are overlaid: the 2014/15 neutrino flare (orange star), and the 9.5-year-averaged emission (dark blue triangle).
}
    \label{fig:PL}
\end{figure*}
\begin{figure*}[t]
    \centering
    \includegraphics[width=1\textwidth]{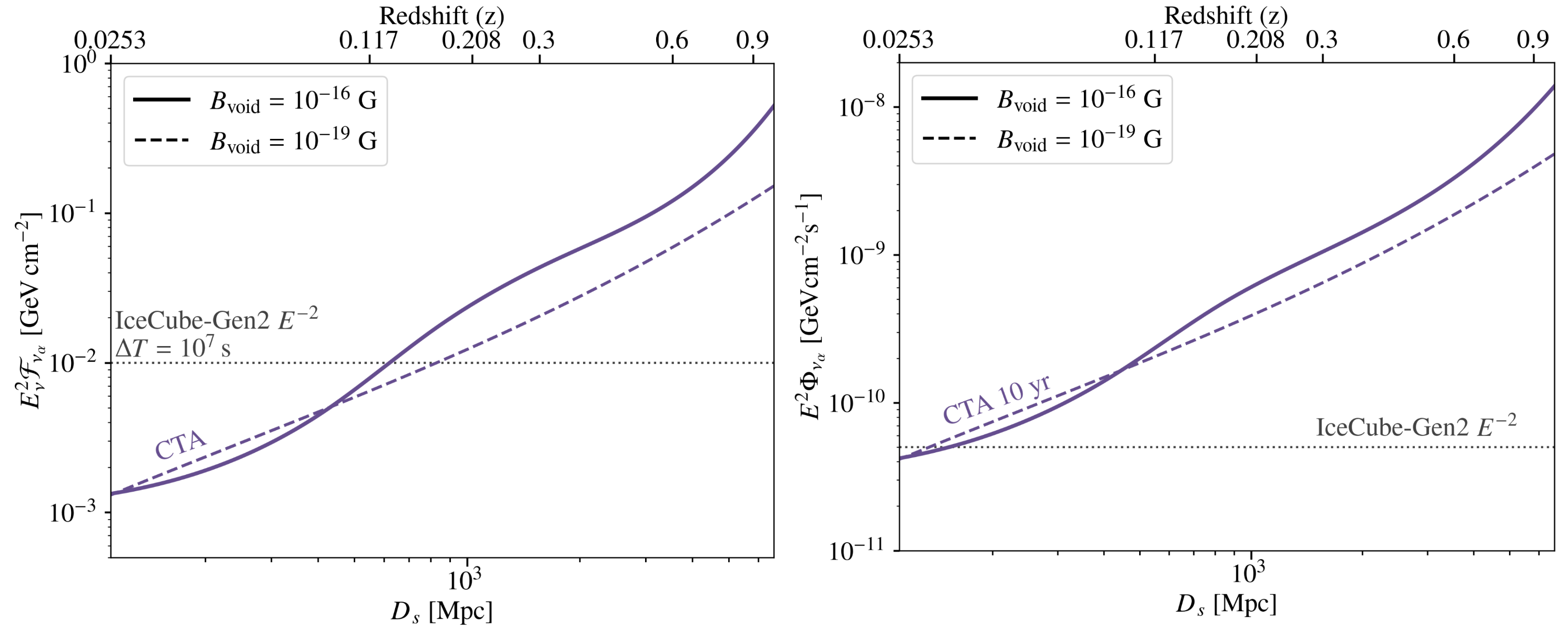}
    \caption{
Required per-flavor neutrino fluence,
$E_\nu^2\mathcal{F}_{\nu_\alpha}$ (left), and flux,
$E_\nu^2\Phi_{\nu_\alpha}$ (right), at Earth as
functions of source luminosity distance $D_s$
(bottom axis) and redshift (top axis), for producing
a detectable $\gamma$-ray cascade counterpart with
the Cherenkov Telescope Array (CTA), assuming an
unbroken $E^{-2}$ power-law neutrino spectrum from
1~TeV to 1~PeV. The left panel corresponds to
transient emission with $\Delta T=10^7$~s, while
the right panel corresponds to steady emission.
The purple curves represent the minimum neutrino
fluence (left) or flux (right) required for the
associated cascade $\gamma$-ray emission to reach
the adopted CTA sensitivity benchmarks, with a
10-year observation assumed for the steady case.
Solid and dashed curves correspond to
$B_{\mathrm{void}}=10^{-16}$~G and
$10^{-19}$~G, respectively. The horizontal
black dotted lines indicate IceCube-Gen2's
point-source fluence (left) and flux (right)
sensitivities, assuming an unbroken $E^{-2}$
spectrum. Regions above both the CTA and
IceCube-Gen2 sensitivity curves define the
multimessenger-detectable region.
}
    \label{fig:PL_CTA}
\end{figure*}
As described in the previous sections, we obtained the $\gamma$-ray flux arriving at Earth for sources embedded in filaments from different redshifts for each IceCube's differential sensitivity.
Then, we determine the neutrino flux threshold corresponding to a detectable $\gamma$-ray counterpart by adjusting the overall normalization of the $\gamma$-ray flux until it intersected the sensitivity of each $\gamma$-ray detector.
Because the injection spectra were tied to IceCube’s differential sensitivity, this normalization directly yields the required neutrino flux in the corresponding energy bin at Earth. 
Repeating this procedure for different redshifts and for both
transient and steady emission scenarios defines the neutrino flux level at which a $\gamma$-ray counterpart becomes detectable.
Considering that IceCube is sensitive to neutrinos above the TeV scale and that $\gamma$-rays above 100\,TeV are attenuated inside the filaments, this work focuses on two energy bins for the observable $\gamma$-ray flux: 1--10\,TeV and 10--100\,TeV, which are shown in Figure~\ref{fig:T_binned} and Figure~\ref{fig:S_binned}. We also examine IceCube's single power-law sensitivity in Figure~\ref{fig:PL}.

Figure~\ref{fig:T_binned} is the transients emission scenario with a flare duration of $\Delta T = 10^7$~s for both energy bins. 
Each panel shows the minimum per-flavor neutrino flux at Earth that is required for the associated cascade $\gamma$-ray emission to be detectable by current $\gamma$-ray observatories as a function of source distance.
We compare the sensitivity threshold of the $\gamma$-ray instruments for 1 year with IceCube's sensitivity for a time window of $\Delta T = 10^7$~s. Solid and dashed curves represent two $B_\text{void}$ strengths, $10^{-16}$~G and $10^{-19}$~G.
A neutrino source with a flux lying above a given curve would produce a $\gamma$-ray cascade that would be detectable by the corresponding instrument; sources above the IceCube line would be detectable in neutrinos. 
Regions above both $\gamma$-ray detector lines and IceCube sensitivity define the “multimessenger horizon” region where a high-energy neutrino event would have an accompanying $\gamma$-ray signal. 

Several important trends emerge from these curves.
First, the required neutrino flux for a detectable $\gamma$-ray counterpart increases steeply with distance for both HAWC and LHAASO.  For the 10--100~TeV bin shown in the right panel in Figure~\ref{fig:T_binned}, this behavior is primarily driven by attenuation of high-energy photons within the filamentary regions, as the EBL density increases with redshift. 
In contrast, the 1--10~TeV bin shown in the left panel exhibits a more complex behavior.  At lower redshifts, a significant fraction of unattenuated primary photons can survive to reach the Earth, and this surviving component dominates the trend. 
At higher redshifts, however, essentially all primary photons are absorbed, and the slope is instead governed by the attenuation of the cascade photons by the EBL and the adiabatic energy loss of the cascade flux.
 Although HAWC and LHAASO can technically detect photons down to $\sim 300$~GeV, this energy lies at the edge of their effective sensitivity. Therefore their capability to detect cascaded flux beyond $z \sim 0.2$ faces a large uncertainty.
As the cascade spectrum shifts to progressively lower energies with increasing propagation distance, HAWC and LHAASO would only be capable of detecting photons near their low-energy cutoff, making the detectability estimates at higher redshifts highly uncertain.

\textit{Fermi}-LAT's curve is nearly flat for 1--10\,TeV bin, shown in the left panel, which indicates that the required neutrino flux for \textit{Fermi}-LAT to detect the $\gamma$-ray cascade does not change much with distance. 

For $B_{\text{void}}=10^{-16}$~G, the required neutrino flux even gradually decreases with redshift. This is because $\gamma$-rays with energies of a few TeV can partially survive over short propagation distances without being fully attenuated by the EBL. 
Thus, the cascade accumulation dominates over redshift losses.
Under the scenario of $B_{\text{void}}=10^{-16}$~G, the time filter removes lower-energy photons. Therefore, as the source redshift increases, the cascade peak shifts toward lower energies, which is closer to \textit{Fermi}-LAT’s optimal sensitivity energy range.
This effect improves detectability despite distance. 
On the other hand, \textit{Fermi}-LAT's sensitivity curves behave differently at higher redshifts for 10--100\,TeV bin. At low redshifts, a consistent build up of the low-energy tail of the cascade spectrum with increasing redshift cancels the attenuation effect inside the filament. At higher redshifts where the cascades are fully developed, the earlier accumulation effect is overtaken by the redshift-dependent attenuation inside the filament, causing a downturn in the cascade spectrum as shown in Figure~\ref{fig:method02}.
Moreover, as expected, the \textit{Fermi}-LAT sensitivity curve is highly sensitive to the strength of $B_{\text{void}}$. Compared to the case with $B_{\text{void}} = 10^{-19}$\,G, a stronger field of $10^{-16}$\,G results in longer time delays, causing more photons below tens of GeV to be filtered out. This significantly reduces the detectability of the cascade by \textit{Fermi}-LAT.

Figure~\ref{fig:S_binned} shows the steady emission scenario for both energy bins. In the stationary-source approximation, propagation-induced time delays do not suppress the time-averaged cascade flux, and therefore no explicit time-delay selection is applied. As a result, the differences between $B_{\rm void}=10^{-16}$ G and $10^{-19}$ G are much smaller than in the transient scenario, with the remaining magnetic-field dependence primarily arising from angular deflections.

We further examine the horizon for the case of power-law spanning over the entire IceCube energy range.
Figure~\ref{fig:PL} illustrates the case where the neutrino spectrum is assumed to follow a single power-law with an index of $E^{-2}$ from 1 TeV to 1 PeV. 
The left panel is the transients scenario, while the right is the steady emission.  Notably, most $\gamma$-rays from 100~TeV to 1~PeV energy ranges are lost in filaments. So the overall behavior of the curves is qualitatively consistent with the results in the binned cases. The thresholds for HAWC and LHAASO rise sharply with distance because the high-energy cascade photons are increasingly absorbed in the filament and the resulting cascade spectrum shifts to lower energies, eventually below the detectors’ effective low-energy limit. 
In contrast, the \textit{Fermi}-LAT thresholds evolve more gradually with redshift. Because of the attenuation inside the filament, the cascade will be dominated by those from the 1-10~TeV bin. The accumulating cascade and shifting of the cascade peak keep the required neutrino flux nearly constant. For the transients scenario, the detectability by \textit{Fermi}-LAT is strongly dependent on the magnetic field: for $B_\text{void} = 10^{-16}$\,G, the time delays filter out lower-energy photons, reducing the observable cascade flux for transient emission compared to the $10^{-19}$\,G case. 

The combined results from Figures~\ref{fig:T_binned}, \ref{fig:S_binned}, and \ref{fig:PL} reveal the multimessenger horizon where we can detect both neutrinos and their $\gamma$-ray counterpart under the assumption that sources are $\gamma$-ray transparent. Hence, the detection or non-detection of one messenger in the presence of the other places strong constraints on the source physics. 

Finally, Figure~\ref{fig:PL_CTA} extends our analysis to future multimessenger observations with CTA and IceCube-Gen2, assuming an unbroken $E^{-2}$ neutrino spectrum. The left panel shows the minimum neutrino fluence required for a detectable $\gamma$-ray counterpart in the transient scenario, while the right panel presents the corresponding neutrino flux for steady emission. At smaller source distances, the CTA threshold lies below the IceCube-Gen2 sensitivity, suggesting that neutrino sources detectable by IceCube-Gen2 could also produce observable $\gamma$-ray counterparts, assuming source transparency. However, the required neutrino fluence or flux increases with distance, eventually exceeding the IceCube-Gen2 sensitivity and making $\gamma$-ray detectability the limiting factor for joint observations. The dependence on the IGMF strength is also evident, with $B_{\mathrm{void}}=10^{-16}$~G generally producing higher thresholds at larger distances than $10^{-19}$~G due to magnetic deflections and, for transient emission, propagation-induced time delays. These results highlight the complementary capabilities of CTA and IceCube-Gen2 for identifying high-energy neutrino sources through their $\gamma$-ray counterparts.

\section{Summary and Outlook}
In this work, we investigated the observational conditions for observation of $\gamma$-ray counterparts of high-energy neutrinos, assuming that the sources are transparent to $\gamma$-rays.
We developed a framework to quantify the minimum neutrino flux required to produce a $\gamma$-ray signal detectable based on the sensitivity of gamma-ray telescopes operating at energies of $\gtrsim$ GeV.
We examined both transient and steady emission and introduced the resulting “multimessenger horizon”, which defines the region in distance and flux that provides a feasible joint observation of gamma rays and neutrinos.

 Defining the multimessenger horizon offers direct insight into the physical environments and emission mechanisms of high-energy neutrino sources. 
Joint detection of neutrinos and gamma rays from a source would offer a unique opportunity to understand the production mechanism and interaction region where mesons are produced. If a source is identified within the horizon projected for the experiments, it would suggest a transparent environment for the production site of neutrinos. On the other hand, an association with a source beyond the horizon would indicate obscuring of the gamma rays at the energies neutrinos are produced. An estimate of the energies at which the gamma rays have cascaded down will bring critical information in understanding the production sites. Additionally, as the horizon depends on the magnitude of IGMF, comparison of the joint observation with the horizon would offer new opportunities for understanding IGMF.
There are a few limitations. We focused on intergalactic electromagnetic cascades governed by the two-photon annihilation and IC scattering. At extremely high energies, synchrotron energy losses are likely to dominate in large-scale structures~\cite{Murase:2011yw,Das:2025zur,Sherman:2025gir}. Second, we assumed that the sources are $\gamma$-ray transparent. Multimessenger studies have suggested that the dominant origin of cosmic neutrinos in the 10~TeV range is $\gamma$-ray hidden, in which our assumption for $\gamma$-ray escape no longer holds. The multimessenger horizons provided in this work cannot directly be applied to such hidden neutrino sources such as AGN coronae~\cite{Murase:2019vdl} and the prompt phase of $\gamma$-ray bursts (except at ultrahigh-energies)~\cite{Razzaque:2004cx,Murase:2009ah}. Third, although we assumed the $p\gamma$ process for demonstrating the multimessenger horizon, the ratio of $\gamma$-rays to neutrinos changes if the Bethe-Heitler process is important and/or neutrinos originate from $\beta$ decay~\cite{Murase:2010va,Murase:2012df,Zhang:2023ewt}.

The framework developed in this work provides a quantitative baseline for interpreting current and future multimessenger observations, and it can be extended in several important directions. As neutrino and $\gamma$-ray detectors continue to improve in sensitivity and energy coverage, the multimessenger horizon will expand, allowing access to a broader population of sources at greater distances and lower flux levels. 
In addition, this framework helps building a predictive, model-independent approach to mapping the neutrino–$\gamma$ connection in the high-energy universe.

\newpage
\appendix

\section{Modifications and Validation of Monte Carlo Simulations}\label{Simulation_Test}
\begin{figure}[t]
    \centering
    \includegraphics[width=1\columnwidth]{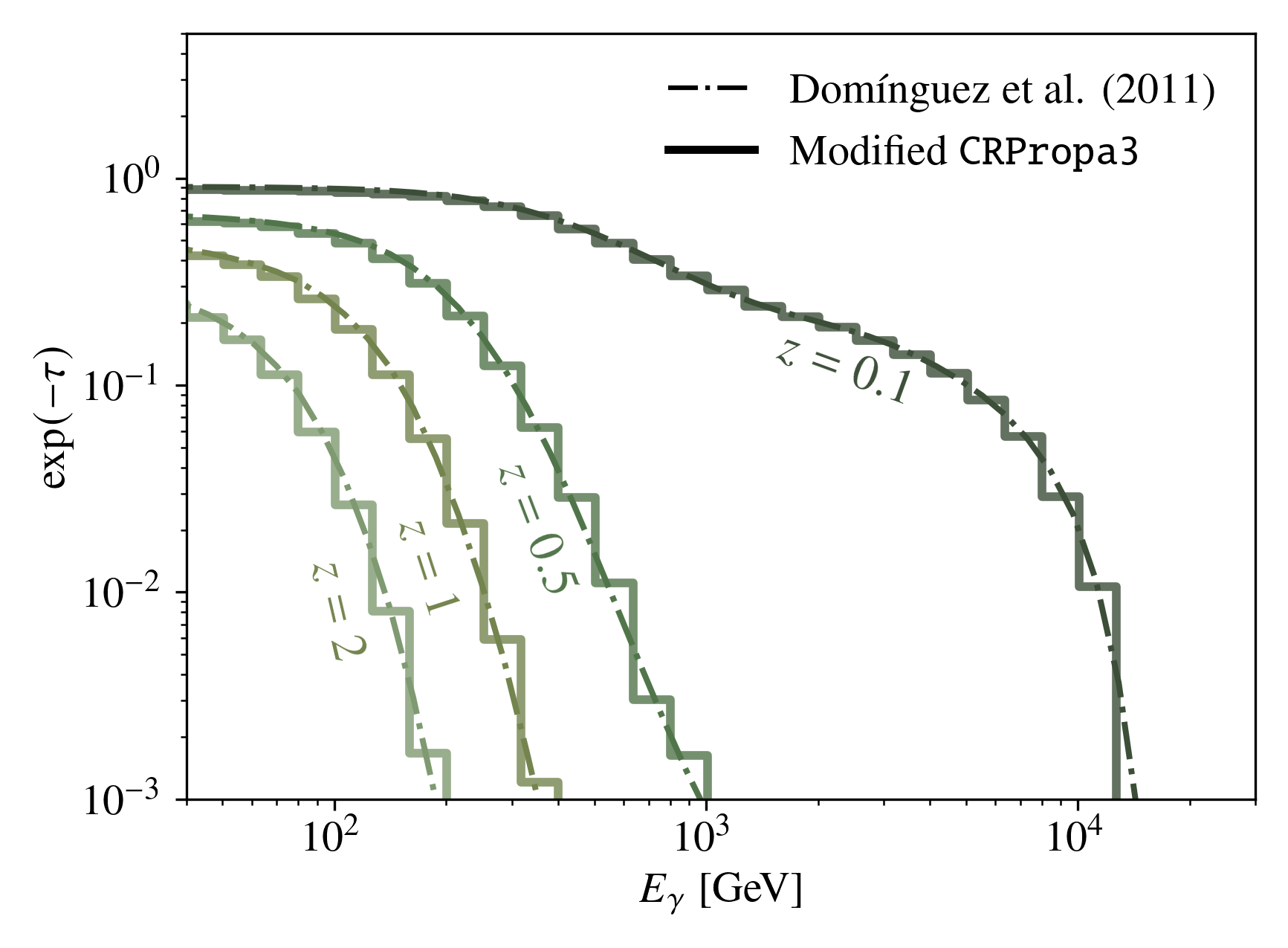}
    \caption{Comparison of the attenuation factor \(\exp(-\tau)\) as a function of observed \(\gamma\)-ray energy \(E_\gamma\) at various redshifts (\(z = 0.1\), 0.5, 1, 2). The dashed curves represent the analytical results based on the Domínguez et al. (2011) EBL model~\cite{Dominguez:2010wxv}, while the solid histograms show the results obtained from the modified \texttt{CRPropa3} simulation using an injected \(\varepsilon_\gamma^{-2}\) spectrum. 
}
    \label{fig: EBL_test}
\end{figure}
\begin{figure}[t]
    \centering
    \includegraphics[width=1\columnwidth]{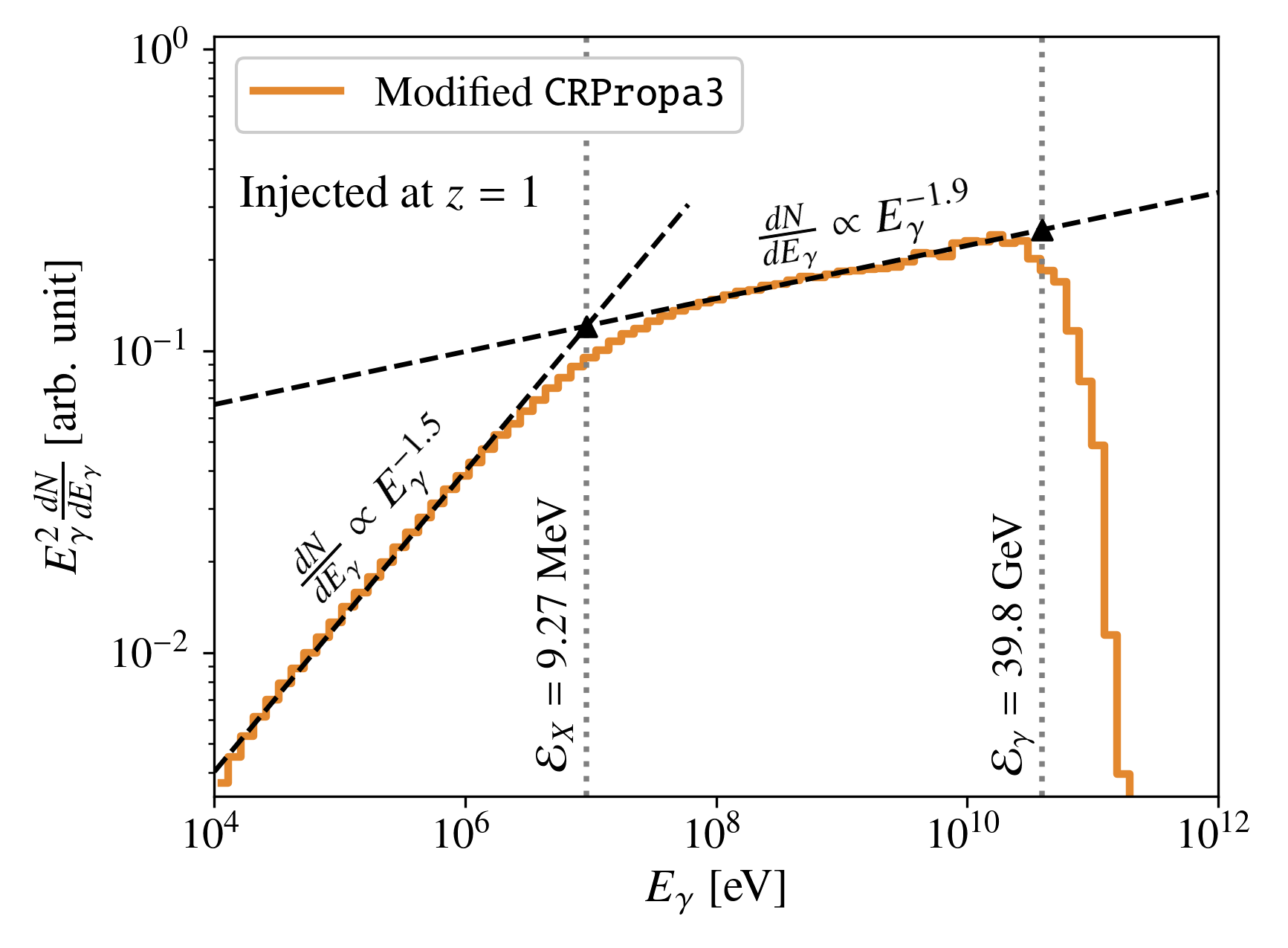}
    \caption{Cascade spectrum produced by injecting \(\gamma\)-ray photons at redshift \( z = 1 \), simulated with the modified \texttt{CRPropa3} (orange solid line). Below the characteristic break energy \( \mathcal{E}_X \approx 9.27\,\mathrm{MeV} \), the spectrum follows a power law of \( dN_\gamma/dE_\gamma \propto E_\gamma^{-1.5} \), while between \( \mathcal{E}_X \) and the cutoff energy \( \mathcal{E}_\gamma \approx 39.8\,\mathrm{GeV} \), it softens to \( dN_\gamma/dE_\gamma \propto E_\gamma^{-1.9} \). The locations of the two turning points match analytical predictions described by Equations~\ref{eq: BP} and \ref{eq: E}.
}
    \label{fig: shape_test}
\end{figure}
In this appendix, we describe the modifications made to the \texttt{CRPropa3} framework and validate the accuracy of our electromagnetic cascade simulations. We first present two major modifications concerning the redshift-dependent interaction rates and the detection geometry used to reconstruct the angular and temporal distributions of cascade photons. We then perform three independent tests to verify the numerical results. First, we compare the simulated $\gamma$-ray attenuation with analytical predictions based on the EBL optical depth. Second, we examine whether the simulated electromagnetic cascades reproduce the expected universal spectral shape. Finally, we investigate the propagation-induced time delays and compare their dependence on photon energy and magnetic field strength with analytical expectations.

\subsection{\texttt{CRPropa3} Modifications}
To improve the accuracy of the electromagnetic cascade simulations, we implemented two major modifications to the original \texttt{CRPropa3} framework. First, the standard implementation of \texttt{CRPropa3} approximates the redshift evolution of electromagnetic interaction rates by applying a scaling factor to precomputed rates, following
\begin{equation}
    \lambda^{-1}(E,z) \simeq (1+z)^2 f(z)\,\lambda_0^{-1}\bigl(E(1+z)\bigr),
\end{equation}
where $\lambda_0^{-1}$ is the interaction rate tabulated at $z=0$, and $f(z)$ represents the redshift-dependent normalization of the background photon field. This approximation does not fully account for the evolution of the EBL spectral shape. Instead, we constructed redshift-dependent interaction-rate tables over a grid of particle energies and redshifts, allowing the rates to be evaluated through interpolation without relying on the approximate scaling prescription. This treatment accounts more accurately for the evolution of the background photon fields, particularly the redshift-dependent spectral distribution of the EBL. Second, we adopted a spherical detection surface centered at the source, with a radius equal to the geometric source--observer distance, to represent the location of Earth. Unlike the planar detection geometry, this spherical surface allows cascade photons to be collected at a fixed distance from the source, independent of their propagation directions. This geometry enables consistent reconstruction of the emission and detection angles relative to the source--observer line of sight. Furthermore, since all detected photons share the same direct source--observer reference distance, the spherical geometry improves the calculation of propagation-induced time delays by avoiding artificial geometric contributions that can arise when the total trajectory length is compared with the perpendicular source-to-plane distance.
\subsection{Validation of the Electromagnetic Cascade Spectrum}

Since this study considers sources at high redshifts up to \( z = 2 \), it is essential to ensure that \texttt{CRPropa3} accurately accounts for the redshift evolution of background photon fields and the adiabatic energy loss of photons. Figure~\ref{fig: EBL_test} presents a comparison between the attenuation factor computed from the simulation and the results from the Domínguez et al. model~\cite{Dominguez:2010wxv}. The optical depth \(\tau(E_\gamma, z)\), which governs the attenuation of high-energy \(\gamma\) rays due to pair production with EBL photons, is typically computed using the standard formalism:
\begin{equation}
\tau(E_\gamma, z) = \int_0^z c\left| \frac{dt'}{dz'} \right| \, dz' \int_{-1}^1 d\mu \, \frac{1 - \mu}{2} \int_{\varepsilon_{\text{th}}}^{\infty} d\varepsilon' \, \sigma_{\gamma\gamma}\, \frac{d n_\text{EBL}}{d\varepsilon'},
\end{equation}
where \( \left| dt'/dz' \right| \) encodes cosmological expansion, \( \varepsilon_{\text{th}} \) is the threshold energy for pair production, \( \sigma_{\gamma\gamma} \) is the pair production cross section, and \( \frac{d n_\text{EBL}}{d\varepsilon'} \) is the proper number density of EBL photons. The attenuation factor is then given by \( \exp(-\tau) \), as implemented in Domínguez et al.’s work. To verify that the simulation reproduces this behavior, we injected \(\gamma\)-ray photons with an \(E_\gamma^{-2}\) spectrum in the modified \texttt{CRPropa3} and numerically extracted the resulting attenuation. As shown in Figure~\ref{fig: EBL_test}, the simulation results are in excellent agreement with the analytical predictions, confirming the correct treatment of  both the redshift-dependent interaction rate and the adiabatic energy loss of photons.

Also, the features of $\gamma$-ray cascades developed in CMB and EBL are also well-studied using an analytic model~\cite{Berezinsky:1975zz,Murase:2012df,Berezinsky:2016feh}. When a high-energy $\gamma$-ray initiates a cascade in the Thomson regime, and the cascade reaches equilibrium by producing multiple generations cascades, the photon spectrum approaches a universal shape that is independent of the initial conditions and depends only on the background photon field. This steady-state cascade spectrum takes the form
\begin{equation}
\label{eq: BP}
\frac{dN}{dE_\gamma} \propto 
\begin{cases}
E_\gamma^{-1.5}, & \text{for } E_\gamma \ll E_{\gamma}^{\rm br} \\
E_\gamma^{-1.9}, & \text{for } E_{\gamma}^{\rm br} \ll E_\gamma \ll E_{\gamma}^{\rm cut}
\end{cases}
\end{equation}
The turning points are defined by cascade break energy $E_{\gamma}^{\rm br}$ and high energy cut-off $E_{\gamma}^{\rm cut}$, which are defined by
\begin{equation}
\label{eq: E}
E_{\gamma}^{\rm cut} = \frac{m_e^2}{\varepsilon^\text{z}_\text{EBL}} \cdot \frac{1}{1+z}, \quad
E_{\gamma}^{\rm br}= \frac{m_e^2 \, \varepsilon_\text{CMB}}{3 \left( \varepsilon^\text{z}_\text{EBL} \right)^2}.
\end{equation}
where \( \varepsilon^\text{z}_\text{EBL} \) is the typical energy of EBL photons at epoch $z$, 
and \( \varepsilon_\text{CMB} \) is the characteristic energy of CMB photons. Below $\mathcal{E}_X$, IC scattering occurs efficiently in the Thomson regime, and photons are no longer energetic enough to pair-produce. Thus, the slope of $-1.5$ in this region arises from balance between IC scattering and energy injection. $E_{\gamma}^{\rm cut}$ is approximately the highest energy at which photons can survive in the cascade before being absorbed. The region between $E_{\gamma}^{\rm br}$ and $E_{\gamma}^{\rm cut}$ is governed by a slope of $-1.9$, due to a balance between pair production and IC scattering of secondaries.

Figure~\ref{fig: shape_test} compares the simulated cascade spectrum with the expected analytical shape for a stationary electromagnetic cascade initiated at redshift \( z = 1 \). The resulting spectrum exhibits a broken power-law behavior, consistent with theoretical predictions, as well as the previous work (see, e.g., Figures~1 and 2 of Refs.~\cite{Murase:2007ar,Murase:2011cy}). A low-energy slope of \( dN_\gamma/dE_\gamma \propto E_\gamma^{-1.5} \) below the characteristic break energy \( E_{\gamma}^{\rm br} \approx 9.27\,\mathrm{MeV} \), and a high-energy slope of \( dN_\gamma/dE_\gamma \propto E_\gamma^{-1.9} \) up to the cutoff energy \( \mathcal{E}_\gamma \approx 39.8\,\mathrm{GeV} \). The agreement confirms that the simulation correctly reproduces the universal cascade shape. 
\subsection{Validation of Propagation-Induced Time Delays}
We further examine the propagation-induced time delays of cascade gamma rays in the weak-field regime. For electrons and positrons undergoing inverse-Compton (IC) cooling, the characteristic magnetic deflection angle can be approximated as
$\delta_e\sim D_{\rm IC}/R_L$, where the IC cooling length and Larmor radius scale as $D_{\rm IC}\propto\varepsilon_e^{-1}$ and $R_L\propto\varepsilon_e/B$, respectively. Consequently, $\delta_e\propto B\varepsilon_e^{-2}$. In the Thomson regime, the characteristic observed energy of photons produced through IC scattering on the CMB satisfies $E_{\gamma,\mathrm{obs}}\propto\varepsilon_e^2$. For small deflections, approximating the relevant geometrical propagation length by the pair-production mean free path $\lambda_{\gamma\gamma}$ gives~\cite{Plaga:1995ins,Murase:2008pe,Takahashi:2008pc, Taylor:2011bn}
\begin{equation}
    \Delta t_B
    \sim \frac{\lambda_{\gamma\gamma}}{2c}\delta_e^2
    \propto B^2 E_{\gamma,\mathrm{obs}}^{-2}
    \lambda_{\gamma\gamma},
    \label{eq:delay_scaling}
\end{equation}
where the approximation assumes a characteristic interaction distance smaller than the source distance. The expression provides an approximate scaling rather than an exact description of multigeneration electromagnetic cascades.

An important consideration is whether magnetic deflections can isotropize the electron--positron pairs before they produce the observed cascade photons. In our simulations, the three-dimensional propagation of charged particles is followed in the turbulent magnetic field without imposing a small-deflection approximation. Thus, large-angle deflections are included in the numerical treatment, subject to the finite maximum trajectory length and other propagation stopping conditions. Nevertheless, significant isotropization is not expected for the electrons responsible for GeV photons through IC scattering on the CMB. In particular, the characteristic deflection angle is approximately
\begin{equation}
    \delta_e \sim 0.11
    \left(\frac{B}{10^{-16}\,\mathrm{G}}\right)
    \left(\frac{E_{\gamma,\mathrm{obs}}}{1\,\mathrm{GeV}}\right)^{-1},
    \label{eq:electron_deflection}
\end{equation}
at low redshift, assuming a magnetic coherence length comparable to or larger than the IC cooling length. Even for our strongest magnetic field, $B_{\rm IGMF}=10^{-16}\,\mathrm{G}$, the characteristic deflection is approximately $0.011$ rad at $E_{\gamma,\mathrm{obs}}=10\,\mathrm{GeV}$ and decreases at higher energies. The corresponding IC cooling lengths are substantially shorter than the maximum propagation length adopted in the simulation. Therefore, the GeV-producing electrons in this regime are expected to remain far from isotropization.

\begin{figure}[t]
    \centering
    \includegraphics[width=0.95\columnwidth]{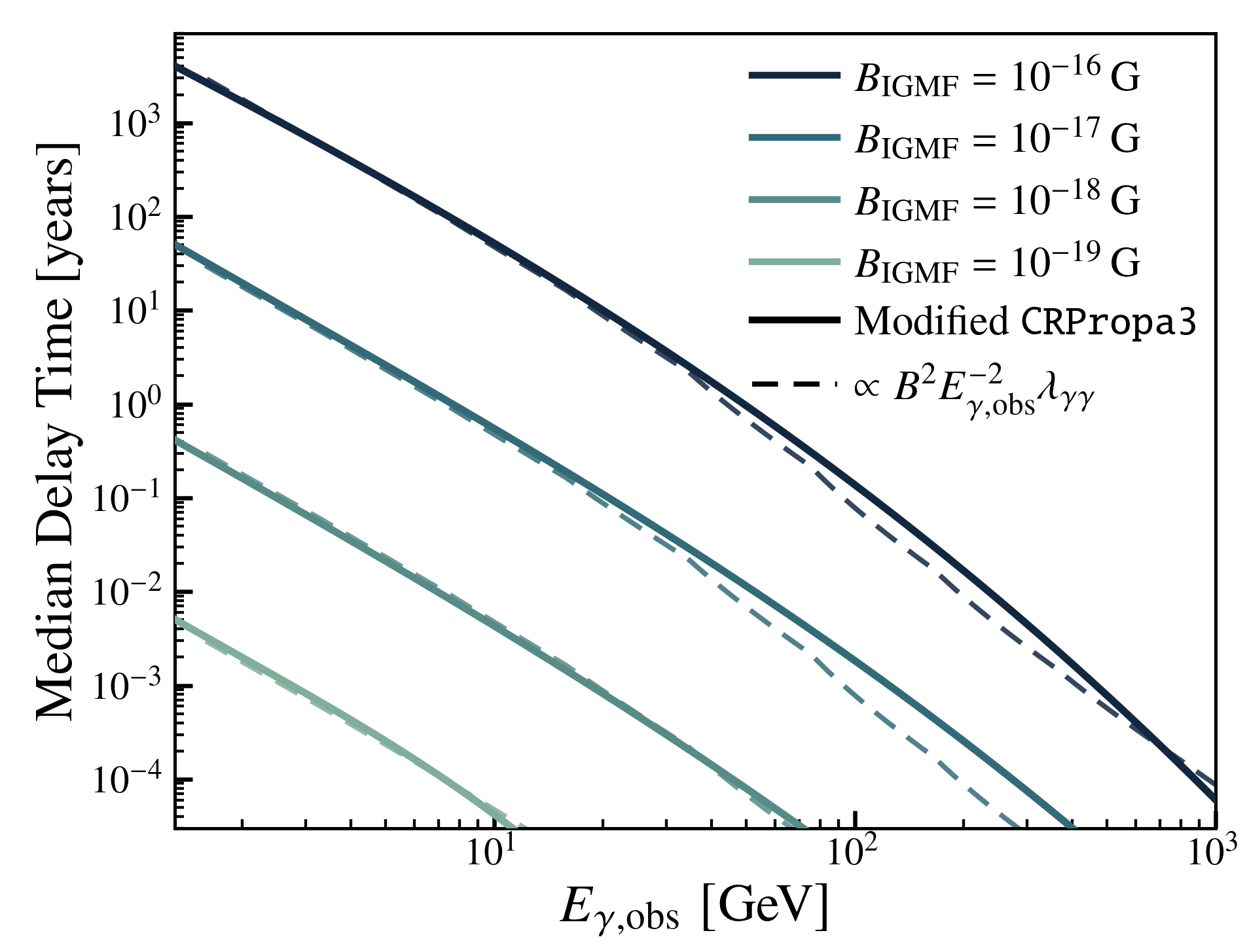}
    \caption{
        Median propagation-induced time delays of cascade gamma rays as a function of observed photon energy for four intergalactic magnetic field strengths. Solid curves represent the modified \texttt{CRPropa3} simulations, with photon trajectories projected onto a source-centered spherical surface of radius equal to the source--observer distance. Dashed curves indicate the approximate weak-field scaling $\Delta t_B\propto B^2E_{\gamma,\mathrm{obs}}^{-2}\lambda_{\gamma\gamma}$.
    }
    \label{fig:delay_validation}
\end{figure}

Figure~\ref{fig:delay_validation} compares the simulated median time delays with the analytical scaling in Eq.~\eqref{eq:delay_scaling} for four IGMF strengths. The results show approximate agreement with the expected magnetic-field and energy dependence, although deviations remain, particularly at higher energies. These differences can arise from the distribution of interaction locations, the energy-dependent electron deflections, and the contributions of multiple cascade generations, which are not fully represented by the analytical approximation. Since the normalization is fixed using the simulation, this comparison tests the scaling behavior rather than the absolute coefficient of the time delay.

\bibliography{ref}

\end{document}